\documentclass[journal]{IEEEtran}

\usepackage{graphicx}
\usepackage{threeparttable}
\usepackage{subfigure}
\usepackage{subcaption} 
\usepackage{caption}
\usepackage{multicol}
\usepackage{booktabs}
\usepackage{epstopdf}
\usepackage{float}
\usepackage{color}
\usepackage{stfloats}
\usepackage{amsmath, amsthm, amssymb, amsfonts}
\usepackage[ruled,linesnumbered]{algorithm2e}
\usepackage{microtype}
\usepackage[noadjust]{cite}
\usepackage{hyperref}
\hypersetup{
  colorlinks=true,
  citecolor=blue,
  linkcolor=magenta,
  urlcolor=blue}
\theoremstyle{definition}

\newtheorem{remark}{Remark}

\begin{document}

\setlength{\lineskiplimit}{0pt}
\setlength{\lineskip}{0pt}
\setlength{\abovedisplayskip}{6pt}  
\setlength{\belowdisplayskip}{6pt}
\setlength{\abovedisplayshortskip}{6pt}
\setlength{\belowdisplayshortskip}{6pt} 

\graphicspath{{images/}}

\title{\textcolor{black}{Breaking the Communication-Accuracy Trade-off: \\ A Sparsified Information Diffusion Framework \\ for Multi-Agent Collaborative Perception}}

\author{Jirong~Zha, Chenyu~Zhao,
        Nan~Zhou,
        Zhenyu~Liu,
        Tao~Sun,
        Bin Zhang,
        Xiaochun~Zhang$^{\ast}$,
        and~Xinlei~Chen$^{\ast}$
\thanks{*Corresponding author.}
\thanks{
J. Zha, C. Zhao, N. Zhou, Z. Liu, and X. Chen are with Shenzhen International Graduate School, Tsinghua University, Shenzhen 518055, China. 
E-mail: zhajirong23@mails.tsinghua.edu.cn, 
zhaocyhi@gmail.com, zhoun24@mails.tsinghua.edu.cn, 
\{zhenyuliu,chen.xinlei\} @sz.tsinghua.edu.cn.
} 
\thanks{T. Sun is with Shenzhen Institute of Artificial Intelligence and Robotics, Shenzhen, China, 518000. E-mail: tsun9tic@gmail.com.}
\thanks{B. Zhang and X. Zhang are with Shenzhen Smart City Technology Development Group, Shenzhen, China, 518000. E-mail: \{zhangbin,irc\}@smartcitysz.com.}
}

\vspace{-3em}

\maketitle

\begin{abstract}

The growing relevance of multi-agent systems has drawn increasing focus on communication-efficient filters for collaborative perception to alleviate the system's communication burden.
While the event-triggered (ET) mechanism can improve communication efficiency in collaborative state estimation, an inevitable trade-off exists between estimation accuracy and communication cost in ET filters.
This paper proposes a fast and accurate ET diffusion-based filter 
for real-time multi-agent collaborative target tracking, aiming to reduce the system's data transmission without compromise in tracking performance. 
The proposed filter achieves improved tracking accuracy, reduced data transmission, and accelerated convergence using an error-minimized ET cubature information filter (CIF) for local estimation, and a correlation-aware diffusion strategy for global fusion.
{The experimental results confirm the scalability of the proposed EDC-CIF algorithm and demonstrate its efficacy in simultaneously reducing estimation error and computation time while significantly enhancing communication efficiency.}

\end{abstract}


\section{Introduction} \label{intro}

Advances in multi-agent collaborative perception, as exemplified by collaborative target tracking for continuous state estimation, have revolutionized collective sensing by harnessing spatial diversity to achieve expansive coverage, high-fidelity estimation, and resilient systemic robustness. \cite{kwon2018sensing, gu2025mr,  cao2012overview, win2011network}.
{Numerous studies have investigated multi-agent collaborative tracking in swarm systems, such as formation control and coordination \cite{abdulghafoor2023multi, liu2017multi, wang2024transformloc, dai2021adaptive}, as well as target search and pursuit \cite{peng2025multi, li2025distributed, hou2023uav, cheng2024multi}.
As unmanned aerial vehicle (UAV) technologies continue to evolve,  collaborative multi-UAV target tracking has found widespread applications in disaster response \cite{wan2024deep, khan2022emerging, wang2025uav, xu2026scalable}, border surveillance \cite{beard2006decentralized, gu2018multiple, zhou2022integrated}, intelligent transportation systems \cite{zhang2023integrated, zhou2021intelligent, sun2024efficient}, and urban monitoring \cite{liu2022smart, stasinchuk2021multi, miao2025multi, zha2026aircopbench}. This is largely attributed to the high mobility, flexibility, and wide-area coverage offered by UAV swarms \cite{du2025survey, javaid2023communication, peng2026resilient}.
Given the necessity of real-time coordination for effective target localization and control, the estimation of the target state in multi-UAV collaborative tracking remains a critical issue \cite{doostmohammadian2021distributed, wang2025aerial, zhang2025cooperative}.
} 

\begin{figure}[t]
    \centering
    \includegraphics[width=1\linewidth]{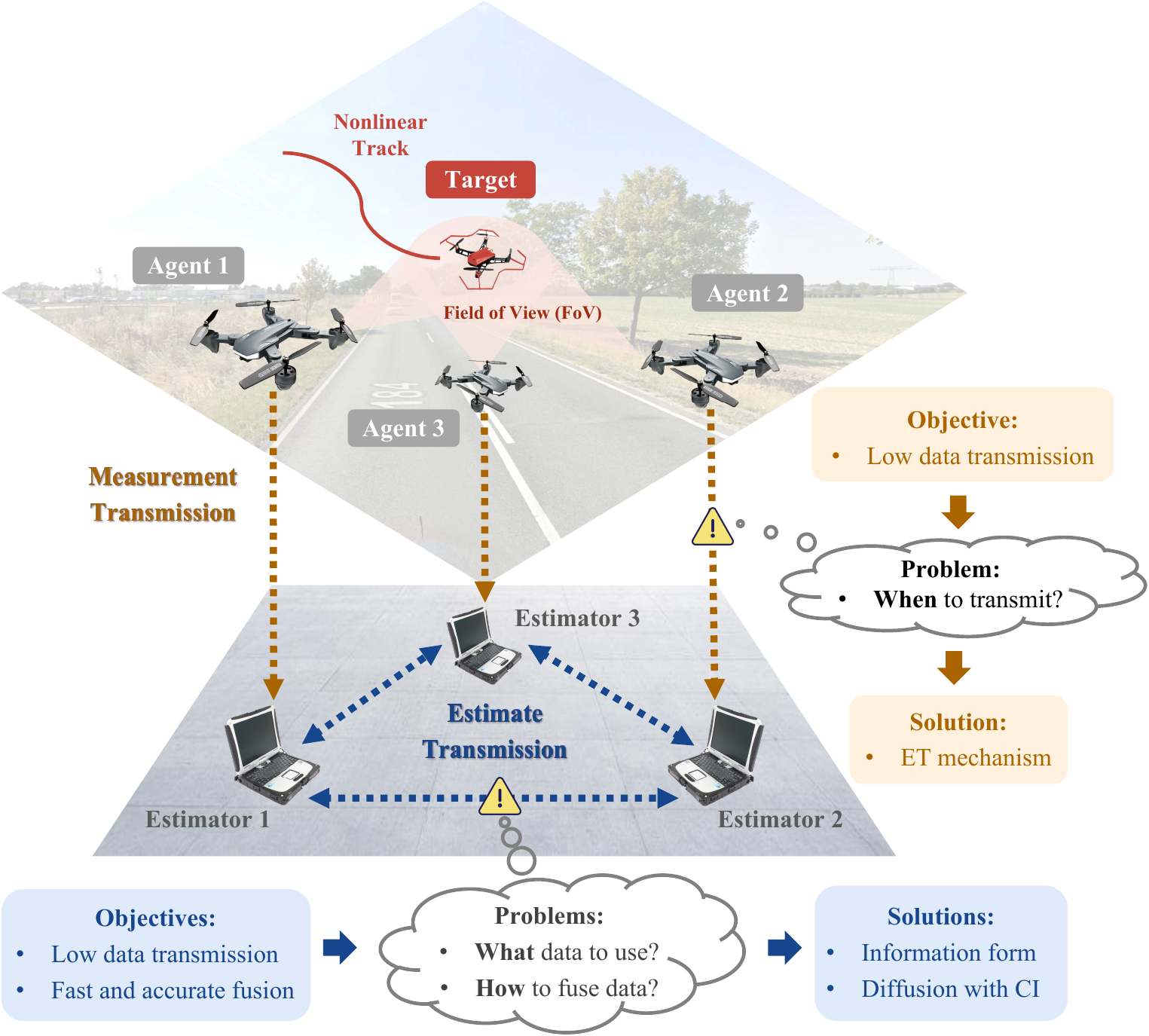}
    \caption{Illustration of collaborative multi-UAV target tracking. \textcolor{black}{A communication-efficient framework is developed for fast and accurate tracking, integrating an event-triggered local filter with a correlation-aware global fusion strategy.}
    }
    \label{system_overview}
    \vspace{-0.5cm} 
\end{figure}

A typical application scenario of  {multi-UAV} collaborative tracking involves a group of drones equipped with effective sensors like cameras serving as multiple agents to collaboratively track a target drone with nonlinear movements,  {estimating the target's state like position and velocity}, as depicted in Fig.~\ref{system_overview}. Each agent generates a local estimate of the target's state at its estimator on the ground station according to measurements collected from its sensor,
and then the estimate is exchanged with neighboring estimators to obtain the target's fused estimated trajectory. Note that the data interaction between estimators is distributed to alleviate communication and computation pressure and to achieve stronger robustness than centralized systems \cite{kamal2013information}.

However, the data transmission burden in distributed collaborative tracking remains an issue due to the frequent communication needs between agents' sensor to estimator and estimator to estimator, which may slow down the system's response speed, occupy more data storage space, and reduce the system's lifetime caused by higher energy consumption \cite{zhou2022integrated}.
 \textcolor{black}{Despite the low sparsity of the signal data exchanged per instance, the primary communication bottleneck in large-scale multi-agent systems is the high frequency, which generates a substantial, cumulative data volume \cite{wang2020learning}.
 Such excessive communication can quickly saturate shared network resources, highlighting the need for communication-efficient collaborative tracking algorithms \cite{berna2004communication}.
 }

This paper aims to tackle distributed collaborative tracking with low data transmission, where multiple sensor-equipped agents and estimators share their local data following specific communication and fusion rules in a lightweight information form. For each agent's sensor and estimator, the main problems are to determine \textit{when to transmit} the local data, \textit{what data to use} for information exchange, and \textit{how to fuse data} with neighboring estimators.

\begin{figure}
    \centering
    \includegraphics[width=1\linewidth]{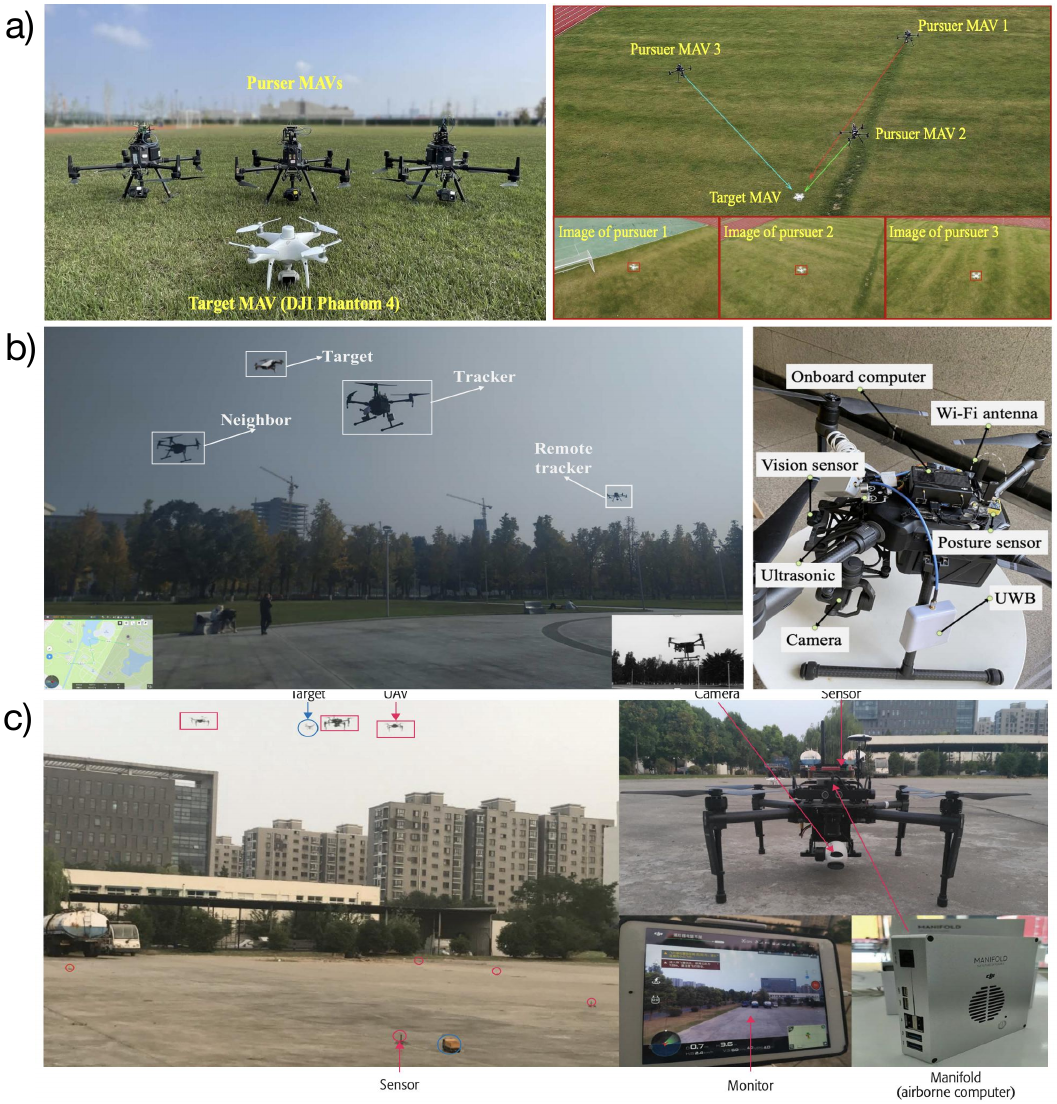}
    \caption{\textcolor{black}{Examples for real-world experiments of collaborative multi-UAV target tracking with three observer UAVs and one target UAV. Image sources: (a) \cite{zheng2025optimal}; (b) \cite{zhou2022integrated}; (c) \cite{gu2018multiple}.}}
    \label{app_ex}
\end{figure}

Existing works on communication-efficient distributed collaborative filters \cite{SECEKF, SECUKF, SECCKF} mainly consider utilizing the event-triggered (ET) mechanism \cite{peng2018survey} to reduce data transmission frequency, estimation pair (i.e., state estimate and error covariance matrix) as transfer data, and consensus \cite{degroot1974reaching} as data fusion approach. However, an inevitable trade-off exists between estimation accuracy and communication cost in conventional ET collaborative filters, i.e., lower data transmission unavoidably leads to a decrease in tracking accuracy \cite{SECUKF}. 
Moreover, the consensus fusion strategy further aggravates the system's communication and computation burden since a sufficient number of iterations is required to make all agents' local estimates reach an agreement. Such limitations of existing ET consensus-based filters prevent them from real-time multi-agent collaborative tracking tasks that demand high accuracy.


This paper aims to propose a distributed collaborative filter that effectively balances the multi-agent system's tracking accuracy, response speed, and communication efficiency.
However, achieving a satisfactory balance of the system's overall tracking performance is non-trivial due to the following two technical challenges:
\begin{itemize}
    \item[1)] Trade-off between estimation accuracy and communication efficiency in local filter (\textbf{C1}). While ET mechanisms
    can be utilized to reduce communication overhead, designing filtering methods tailored to ET is challenging due to the uncertainty in the time steps at which measurements are updated. 
    \item[2)] Unknown sensor correlations in global fusion (\textbf{C2}). As sensor correlation is typically unknown, determining the optimal fusion approach is challenging. The estimation accuracy will suffer if the sensor data are assumed to be independent. Therefore, a more sophisticated and accurate fusion approach needs to be developed.
\end{itemize}

Motivated by the above discussions, we develop an event-triggered diffusion-based cubature information filter with covariance intersection (EDC-CIF) for fast and accurate collaborative tracking with low data transmission, where an event-triggered cubature information filter (ET-CIF) serves as the local filter at each agent and diffusion fusion with covariance intersection (CI) acts as the global fusion strategy. 

To address \textbf{C1}, we design an \textit{error-minimized ET filter} by minimizing ET-CIF's error covariance matrix. This filter effectively reduces the measurement transmission frequency without notable compromise in estimation accuracy and provides information pair (i.e., information vector and matrix) as output for lightweight data fusion.

To address \textbf{C2}, we propose a \textit{correlation-aware diffusion fusion strategy} considering unknown sensor correlations. This fusion technique with CI converges faster and obtains more accurate results by weighting iterative data according to the error covariance matrix compared to consensus.

 {Experimental results demonstrate that the proposed EDC-CIF algorithm effectively reduces estimation error and computation time while enhancing communication efficiency.}
The contributions of this paper are in four aspects:

\begin{itemize}
    \item We propose a novel communication-efficient collaborative filter, EDC-CIF, to alleviate the system's communication pressure, prevent numerous fusion iterations, and offset accuracy loss caused by ET mechanisms.
    
    
    

    \item For local estimation, we design an \textit{error-minimized ET filter} with minimal accuracy loss. For global fusion, we design a \textit{correlation-aware diffusion fusion strategy} with CI to deal with unknown sensor correlations.
    
    
    \item We analyze the proposed filter's stability by proving the stochastic boundedness of its estimation error in the mean square and give a corresponding theorem. 

    \item \textcolor{black}{We carry out both numerical simulations and real-world experiments\footnote{\textcolor{black}{Several representative real-world experiments on multi-UAV collaborative target tracking are illustrated in Fig.~\ref{app_ex}. These examples highlight typical experimental configurations adopted in existing studies, where a small number of observer UAVs and target UAV are used to validate cooperative sensing and tracking capabilities in practical environments.}} to validate our algorithm's feasibility for collaborative tracking and show its superiority in accuracy, fastness, and communication efficiency.}
    
\end{itemize}

The rest of this paper is organized as follows. Section~\ref{related work} presents the related work. Section~\ref{overview} gives an overview of the collaborative tracking system, including the architecture, preliminaries, and primary modules. Section~\ref{filtering} develops the communication-efficient collaborative filter, EDC-CIF. 
\textcolor{black}{
Eventually, Section~\ref{evaluation} conducts numerical simulations and real-world experiments to evaluate the algorithm's tracking performance.}


\textit{Notation}. 
Let $\mathbb{R}^n$ and $\mathbb{R}^{n \times m}$ denote a real $n$-dimensional Euclidean vector space and a real $n \times m$ matrix space, respectively. $\mathbb{E}\{\cdot\}$ stands for the expectation operation, and $\Vert\cdot\Vert$ is the Euclidean norm of a vector. For a matrix $\boldsymbol{A}$, $\boldsymbol{A}^{-1}$ and $\boldsymbol{A}^{\rm{T}}$ represent, respectively, its inverse and transpose. $\text{tr}(\boldsymbol{A})$ means the trace of $\boldsymbol{A}$. The notation $x\sim\mathcal{N}(\mu,\sigma^2)$ denotes that random variable $x$ follows the Gaussian distribution with mean $\mu$ and variance $\sigma^2$. $\boldsymbol{I}_{N}$ indicates the $N \times N$ identity matrix. $\text{col}(\cdot)$ represents the operation to aggregate all the column vectors into a single column vector. Denote by $\text{diag}\{p_1,p_2,\dots,p_n\}$ the block diagonal matrix with $p_i,i\in\{1,\dots,n\}$ on the diagonal.
\section{Related Work} \label{related work}
This section presents related work on various ET mechanisms, communication-efficient local filters based on ET mechanisms, data fusion techniques for collaborative filters, and covariance intersection fusion.

\begin{figure*}
    \centering
\includegraphics[width=0.7\linewidth]{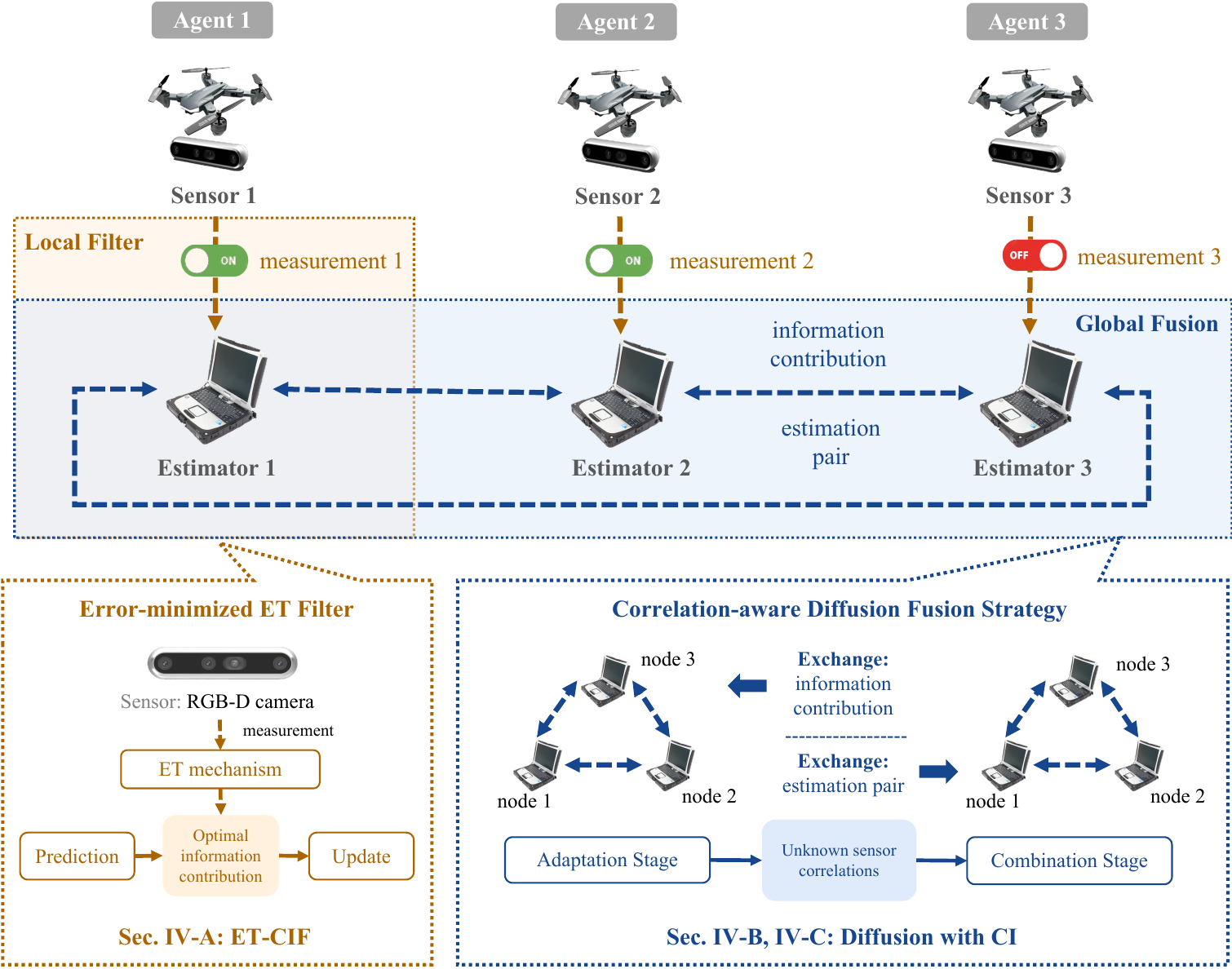}
    \caption{Architecture of the multi-agent collaborative tracking system.
    To ensure accurate nonlinear estimation with low data transmission, each sensor measurement is transmitted to the local CIF via the ET mechanism based on its information contribution. Local estimates are then shared through a diffusion strategy for efficient fusion, and the final result is computed using CI to account for unknown inter-sensor correlations.
    }
    \label{architecture}
\end{figure*}

\vspace{-0.35cm}
\subsection{Event-Triggered Mechanisms}
To reduce the frequency of data transmission, numerous related works \cite{rezaei2022event, SECCKF, SECUKF} have considered ET communication mechanisms \cite{peng2018survey}, which rely on specific threshold conditions to determine whether to transfer current data according to the discrepancy of data at different time instants. Given different triggering conditions, ET mechanisms can be categorized as the ones based on the Innovation-Level condition \cite{han2015stochastic} and the ones following the Send-on-Delta (SoD) schedule \cite{miskowicz2006send}. The latter is more commonly utilized since it considers the difference between the current measurement and the latest transmitted measurement instead of the predicted one, indicating there is no extra computation or prediction error, thus is simpler and faster. Furthermore, depending on the types of triggering thresholds, ET mechanisms can be further classified as static ones \cite{hu2015consensus} with predefined fixed thresholds and dynamic ones \cite{ge2021dynamic} with adjustable changing thresholds. Generally, dynamic schemes are more efficient in communication resource saving than static solutions \cite{song2020distributed}. 
Moreover, by considering diverse triggering links, triggers can occur in the sensor-to-estimator channels \cite{zhang2017distributed} with measurements serving as the trigger variable and estimator-to-estimator channels \cite{li2016event} where the estimated states are the variables to be determined whether to transmit or not. 
\textcolor{black}{Recently, the paradigm of ET mechanisms has been extended to the hardware level with the emergence of event cameras \cite{wang2026event, wang2025ultra}, which skip redundant static scenes and only transmit pixel-level intensity changes, effectively reducing data redundancy.}

\vspace{-0.4cm}
\subsection{Communication-Efficient Local Filters}
Based on ET mechanisms, various communication-efficient filters for state estimation have been developed. For instance, the ET Kalman filter (ET-KF) \cite{battistelli2018distributed} is a technique for linear systems containing Gaussian noises. To achieve nonlinear tracking, an ET extended Kalman filter (ET-EKF) \cite{rezaei2022event} is proposed. For systems with high nonlinearities, the ET unscented Kalman filter (ET-UKF) \cite{SECUKF} is designed with more accurate estimation results compared with ET-EKF. Moreover, the ET cubature Kalman filter (ET-CKF) \cite{SECCKF} is introduced to enhance computational stability. However, few studies have focused on the ET cubature information filter (ET-CIF), which is an algebraic equivalence of ET-CKF and is more robust to numerically sensitive operations. To the best of our knowledge, only ET-CIF with direct trigger operations \cite{tan2018distributed, lin2022distributed} has been considered, but not the optimal scheme that minimizes the error covariance matrix. 

\vspace{-0.3cm}
\subsection{Data Fusion for Collaborative Filters} 
As a crucial part of collaborative target tracking, the way of data fusion significantly affects the estimation performance \cite{zha2023privacy}. Compared with centralized solutions, distributed state estimation (DSE) \cite{marano2008distributed, marano2006quantizer} based on peer-to-peer interaction among sensor nodes has more development potential due to its high scalability, strong robustness, and low resource consumption \cite{kamal2013information}. Currently, there exist two main branches of fusion strategies for DSE, including consensus-based methods \cite{degroot1974reaching} and diffusion-based schemes \cite{cattivelli2010diffusion}. The former is dedicated to reaching an agreement between all nodes' estimates, while the latter is devoted to reducing the estimation error at each node \cite{liang2023event}.
It has been shown that diffusion networks converge faster with a lower mean-square deviation than consensus networks in real-time fusion \cite{tu2012diffusion}, thus making it gain growing attention. Existing research has covered diffusion-based KF \cite{vahidpour2019partial}, EKF \cite{cattivelli2010distributed}, and UKF \cite{chen2021distributed}. However, rare research has studied diffusion-based CKF and CIF, let alone the corresponding algorithms under ET mechanisms. An ET diffusion-based CKF is developed in \cite{liang2023event}, but its transmitted data remains redundant due to the iterative form of CKF. 
Therefore, a more appropriate form of the nonlinear filter is needed for the diffusion framework.

\vspace{-0.35cm}
\subsection{Covariance Intersection}
During the data fusion process, the collaborative filter's accuracy can be enhanced further by considering sensor nodes' relevance. Since the cross-correlations between estimates from different nodes are usually unknown, it may deteriorate the filter's estimation performance if the fused results are generated by a convex combination of local estimates regardless of cross-covariance matrices \cite{chen2021distributed}. To address such issues, some research resorts to the covariance intersection (CI) fusion technique \cite{CI} and presents a series of CI-based collaborative filters \cite{gao2023robust, sun2023inverse, hu2011diffusion, hu2023distributed}. Nonetheless, the majority of CI techniques currently in use are limited to time-triggered communication protocols \cite{liang2023event}. Thus, this paper is also committed to developing an ET diffusion-based CIF algorithm considering CI fusion.

\section{System Overview} \label{overview}
This section illustrates the architecture of the multi-agent collaborative tracking system in Fig.~\ref{architecture}. As seen, our proposed communication-efficient collaborative filter mainly consists of two parts, the local filter at each observing agent for nonlinear target state estimation, and the global fusion for final estimated results.
An observation system consisting of multiple agents equipped with sensors like RGB-D cameras or {LiDARs} and estimators that can interact with each other can be regarded as a sensor network, and each agent acts as a sensor node. Therefore, a multi-agent collaborative target tracking problem can be algorithmically modeled as a multi-sensor collaborative state estimation issue. 


\vspace{-0.25cm}
\subsection{Problem Formulation}
Before introducing the technical scheme of the multi-agent collaborative filter, we build the mathematical models of the communication topology between multi-observation nodes, the moving target's dynamic function, and each sensor's observation equation.

\subsubsection{Collaborative Graph} 
We utilize the mathematical concept of graphs to describe the sensor nodes' communication relationship. Consider a sensor network with $N$ nodes as an undirected graph $\boldsymbol{\mathcal{G}}=\left\{\boldsymbol{\mathcal{V}},\boldsymbol{\mathcal{E}}\right\}$, where $\boldsymbol{\mathcal V}=\left\{\nu_1,\nu_2,...,\nu_N\right\}$ represents the vertex set, and $\boldsymbol{\mathcal E}\subset \boldsymbol{\mathcal V}\times\boldsymbol{\mathcal V}$ indicates the edge set. If there exists an edge between $\nu_i$ and $\nu_j$, the two nodes can exchange information, and $\nu_j$ is called the neighboring node of $\nu_i$. Given node $\nu_i$'s neighbor set as $\boldsymbol{\mathcal N}_i=\left\{\nu_j\in \boldsymbol{\mathcal{V}}\backslash\left\{\nu_i\right\}|(\nu_i,\nu_j)\in \boldsymbol{\mathcal{E}}\right\}$, its cardinality $|\boldsymbol{\mathcal N}_i|$ signifies the number of $\nu_i$'s adjacent nodes. To ensure the information traversal over the entire network, the undirected graph is required to be connected. In particular, if there exists at least one path between any pair of nodes in $\boldsymbol{\mathcal{V}}$, the $N$-sensor graph $\boldsymbol{\mathcal{G}}$ is connected.

\subsubsection{Target Tracking Model}
{In the collaborative tracking problem, the noisy measurements from each sensor constitute the input, while the estimated target state serves as the output.}
We consider nonlinear system dynamics and observations in discrete-time scenarios \cite{zha2026dimm}. The mathematical representation of an $N$-sensor target tracking system follows
\begin{equation}
	\left\{
	\begin{aligned}
		\boldsymbol{x}_k&=f\left( \boldsymbol{x}_{k-1}\right) +\boldsymbol{w}_{k-1},  \\
		\boldsymbol{z}_k^i&=h^i\left( \boldsymbol{x}_k\right) +\boldsymbol{v}_k^i, \ \nu_i\in\boldsymbol{\mathcal{V}},
	\end{aligned}
	\right.
	\label{eq track_model}
\end{equation}
where $\boldsymbol{x}_k\in\mathbb{R}^n$ stands for target's state vector at time instant $k$, and $\boldsymbol{z}_k^i\in\mathbb{R}^m$ refers to the $i$th sensor's measurement vector relative to the target state at the same time instant. Function $f(\cdot): \mathbb{R}^n\rightarrow\mathbb{R}^n$ specifies the nonlinear state transition model of the moving target, and $h^i(\cdot): \mathbb{R}^n\rightarrow\mathbb{R}^m$ denotes the nonlinear measurement model of sensor $\nu_i$. The target's movement process noise $\boldsymbol{w}_{k-1}\in\mathbb{R}^n$ and each sensor's measurement noise $\boldsymbol{v}_k^i\in\mathbb{R}^m$ are both white Gaussian following $\boldsymbol{w}_{k-1}\sim\mathcal{N}(0,\boldsymbol{Q}_{k-1})$ and $\boldsymbol{v}_k^i\sim\mathcal{N}(0,\boldsymbol{R}_k^i)$, respectively, where $\boldsymbol{Q}_{k-1}\in\mathbb{R}^{n\times n}$ and $\boldsymbol{R}_k^i\in\mathbb{R}^{m\times m}$ are the corresponding covariance matrices of the noises.

\vspace{-0.35cm}
\subsection{Local Filter}
The local filter at each observation node is configured as a CIF with static ET mechanism to reduce measurement transmission overhead from sensor to estimator. A brief introduction to CIF and ET mechanism is given below.

\subsubsection{Cubature Information Filter} \label{CIF}
First, CKF \cite{arasaratnam2009cubature} is introduced as the basis of CIF. 
{At each sensor node $\nu_i\in\boldsymbol{\mathcal{V}}$, the local CKF procedure is illustrated in Appendix A\footnote{The appendix of this paper is available at:
\url{https://github.com/zhajirong/TMC-EDC-CIF/blob/main/appendix.pdf}.}.
As an information equivalence of CKF, 
CIF \cite{pakki2011cubature} computes the information vector and matrix as iteration variables and updates based on the information contributions, further enhancing the filter's numerical stability. 
For node $\nu_i$ at time instant $k$ in the prediction phase, the information vector $\boldsymbol{\hat y}_{k\vert k-1}^i$ and information matrix $\boldsymbol{Y}_{k\vert k-1}^i$ are substituted in CIF for the prior estimate $\boldsymbol{\hat x}_{k\vert k-1}^{i}$ and error covariance matrix $\boldsymbol{P}_{k\vert k-1}^i$ in CKF, respectively,  which are calculated by
\begin{equation}
	\begin{aligned}
		\boldsymbol{\hat y}_{k|k-1}^{i}=(\boldsymbol{P}_{k|k-1}^{i})^{-1}\boldsymbol{\hat x}_{k|k-1}^{i}, \ 
		\boldsymbol{Y}_{k|k-1}^{i}=(\boldsymbol{P}_{k|k-1}^{i})^{-1}.
	\end{aligned}
	\label{eq prior_info}
\end{equation}
As for the update phase of CIF, the information contribution $\boldsymbol{i}_k^{i}$ and correlation information matrix $\boldsymbol{I}_k^{i}$ follow
\begin{equation}
	\left\{
	\begin{aligned}
		\boldsymbol{i}_k^{i}=&\boldsymbol{Y}^i_{k\vert k-1}\boldsymbol{P}^i_{x_k,z_k}( \boldsymbol{D}_k^{i})^{-1} \\
        &\times\left[\Delta\boldsymbol{z}_k^{i}+(\boldsymbol{P}^i_{x_k,z_k})^{\rm T}\boldsymbol{Y}^i_{k\vert k-1}\boldsymbol{\hat x}_{k|k-1}^{i}\right] ,  \\
		\boldsymbol{I}_k^{i}=&\boldsymbol{Y}^i_{k\vert k-1}\boldsymbol{P}^i_{x_k,z_k}( \boldsymbol{D}_k^{i})^{-1}(\boldsymbol{P}^i_{x_k,z_k})^{\rm T}\boldsymbol{Y}^i_{k\vert k-1},
	\end{aligned}
	\right.
	\label{eq i and I}
\end{equation}
where
\begin{equation*}
    \begin{aligned}
        \boldsymbol{D}_k^{i}=&\boldsymbol{P}_{z_k,z_k}^i-(\boldsymbol{P}^i_{x_k,z_k})^{\rm T}\boldsymbol{Y}^i_{k\vert k-1}\boldsymbol{P}^i_{x_k,z_k},
	\label{eq D}
    \end{aligned}
\end{equation*}
and $\Delta\boldsymbol{z}_k^{i}=\boldsymbol{z}_k^{i}\!-\boldsymbol{\hat z}_{k\vert k-1}^{i}$. Then, the information pair including both the information vector and matrix is updated as
\begin{equation}
	\begin{aligned}
		\boldsymbol{\hat y}_{k}^{i}=\boldsymbol{\hat y}_{k|k-1}^{i}+\boldsymbol{i}_k^{i}, \ 
		\boldsymbol{Y}_{k}^{i}=\boldsymbol{Y}_{k|k-1}^{i}+\boldsymbol{I}_k^{i},
	\end{aligned}
	\label{eq posterior_info}
\end{equation}
from which, one can eventually obtain the posterior estimate and error covariance matrix as given in \eqref{eq posterior_xp} for the next time instance.
\begin{equation}
	\boldsymbol{\hat x}_{k}^{i}=(\boldsymbol{Y}_{k}^{i})^{-1}\boldsymbol{\hat y}_{k}^{i},\ \boldsymbol{P}_{k}^{i}=(\boldsymbol{Y}_{k}^{i})^{-1}.
	\label{eq posterior_xp}
\end{equation}
The detailed derivation of~\eqref{eq posterior_info} is
provided in Appendix B. 
\vspace{0.01cm}


\subsubsection{Event-Triggered Mechanism} \label{ET}
To reduce redundant transmissions in collaborative tracking, the ET mechanism is integrated into the CIF update phase. It triggers data transmission only when the measurement change exceeds a threshold, filtering out low-impact observations and alleviating communication load.
For static ET schemes, the trigger threshold $\delta^i>0\ (\nu_i\in\boldsymbol{\mathcal V})$ is a fixed predetermined value. Define the trigger flag of node $\nu_i$ at instant $k$ as
\begin{equation}
	\gamma_k^i=\left\{
	\begin{aligned}
		1 & ,\ \text{if}\ (\boldsymbol{z}_k^i-\boldsymbol{z}_{\tau^i_{k-1}}^i)^{\rm T}(\boldsymbol{z}_k^i-\boldsymbol{z}_{\tau^i_{k-1}}^i)>\delta^i,  \\
		0 & ,\ \text{otherwise},
	\end{aligned}
	\right.
	\label{eq gamma}
\end{equation}
where 
$\boldsymbol{z}_{\tau^i_{k-1}}^i$ stands for the last transmitted measurement and updates from
\begin{equation}
    \boldsymbol{z}_{\tau^i_{k}}^i = \gamma_k^i\boldsymbol{z}_k^i+(1-\gamma_k^i)\boldsymbol{z}_{\tau^i_{k-1}}^i.
    \label{eq z_tau}
\end{equation}
Correspondingly, the current measurement transmitted time instant $\tau_k^i$ of sensor $\nu_i$ follows
\begin{equation*}
    \tau_k^i=\left\{
	\begin{aligned}
       & k,\ \ \ \ \ \text{if}\ \gamma_k^i=1,  \\
	  & \tau^i_{k-1}  ,\ \text{if}\ \gamma_k^i=0.
    \end{aligned}
	\right.
    \label{eq tau}
\end{equation*}
Eq.~\eqref{eq z_tau} indicates that the data transmission is allowed only if the trigger flag satisfies $\gamma_k^i=1$, otherwise, the actual utilized measurement from sensor $\nu_i$ remains unchanged. As depicted in Fig.~\ref{architecture}, the local filter with an ET mechanism at each node is equivalent to adding a trigger switch to the measurement transmission link between the sensor and estimator. 

\vspace{-0.3cm}
\subsection{Global Fusion} \label{Global Fusion}
As for distributed data fusion, the diffusion strategy \cite{cattivelli2010diffusion} is utilized to obtain the global estimate through mutual communication among adjacent nodes.
Diffusion-based fusion consists mainly of two stages, as illustrated in Fig.~\ref{architecture}:
\begin{itemize}
	\item \textbf{Adaptation Stage:} Each node shares the information contribution and correlation information matrix with its neighboring nodes and generates its local posterior state estimate according to the interacted data in this stage.
	\item \textbf{Combination Stage:} Each node exchanges the local estimate and error covariance computed in the adaptation stage with its neighbors through covariance intersection fusion and outputs the global fused estimation pair. 
\end{itemize} 

\begin{remark}
    As shown in \cite{cattivelli2010distributed}, diffusion-based EKF, UKF, and CKF incur high overhead as sensor nodes and measurement dimensions increase. To avoid communication overload, we adopt the equivalent information form of CKF (CIF), which transmits fixed-size information matrices, as detailed in Sec.~\ref{ED-CIF}.
\end{remark}
\section{Collaborative Filter} \label{filtering}

This section introduces the proposed 
EDC-CIF algorithm
by designing a communication-efficient local filter and deducing the diffusion-based collaborative filter with appropriate adaptation and combination rules. For ease of understanding, Fig.~\ref{fig framework} depicts the framework of EDC-CIF.

\begin{figure*}[h!]\color{blue}
	\centering
	\includegraphics[scale=0.5]{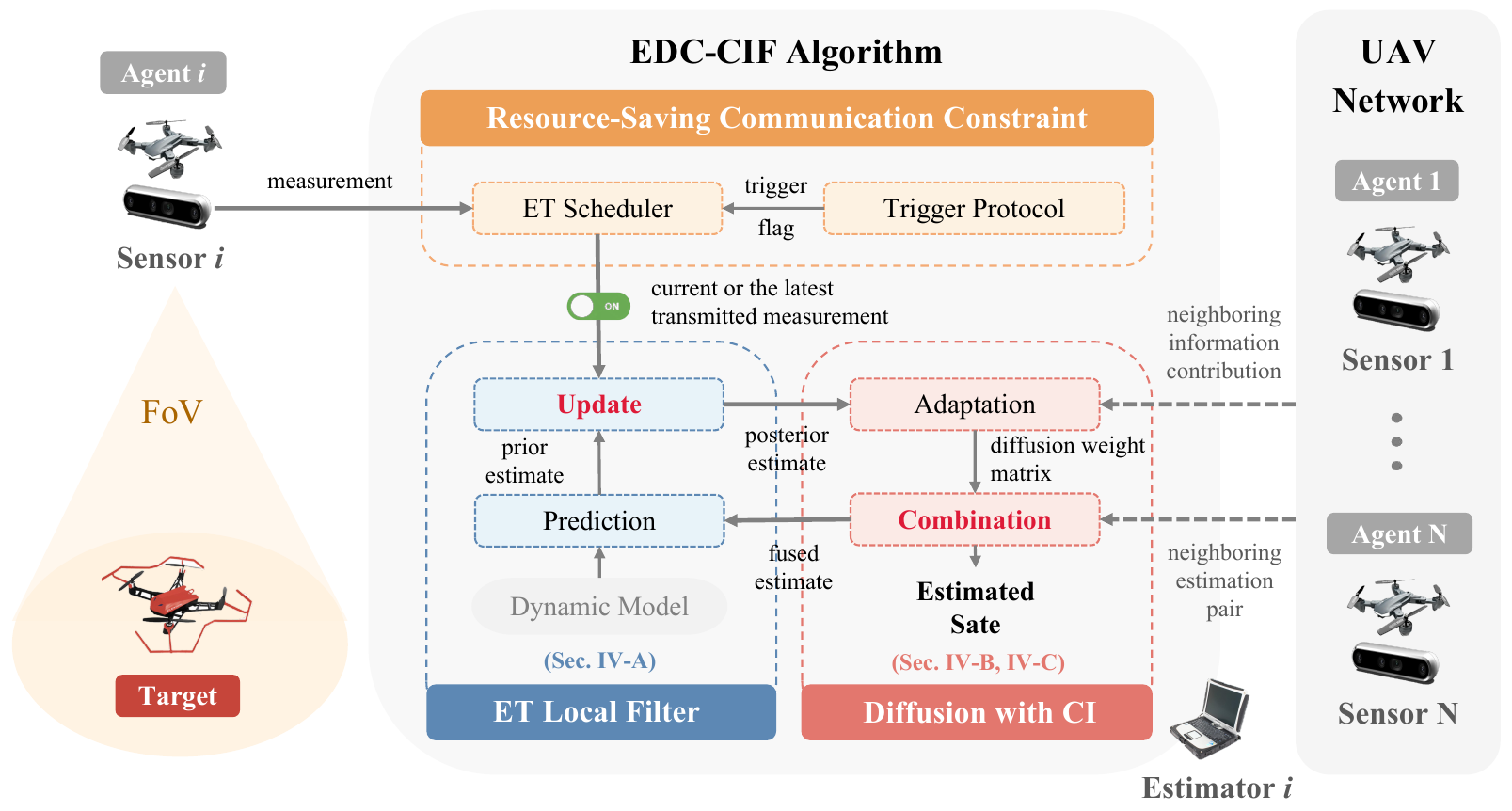}
	\caption{Framework of the EDC-CIF algorithm. The innovation of this algorithm mainly lies in the state update of the local ET nonlinear filter with optimal information contribution, the information form of data utilized for exchange in diffusion fusion, and a novel diffusion combination strategy considering unknown sensor correlations. The algorithm's innovative parts have been highlighted in red for ease of noting.}
	\label{fig framework}
  \vspace{-0.5cm} 
\end{figure*}

\vspace{-0.3cm}
\subsection{Event-Triggered Filter} \label{ET-CIF}
The basic ET-CKF is first designed, then its information-equivalent form, i.e., the ET-CIF, is further proposed. 

\subsubsection{Event-Triggered CKF}
Applying the ET mechanism to the measurement update of local CKF at each time instant, one can obtain the posterior estimate as
\begin{equation}
	\boldsymbol{\hat x}_{k}^i=\boldsymbol{\hat x}_{k\vert k-1}^i+\boldsymbol{\bar K}_k^i(\boldsymbol{z}_{\tau_{k}^i}^i-\boldsymbol{\hat z}_{k\vert k-1}^i),
	\label{eq ET-CKF_upx}
\end{equation} 
where $\boldsymbol{\bar K}_k^i=\gamma_k^i\boldsymbol{K}_k^i+(1-\gamma_k^i)\boldsymbol{M}_k^i$ is the updated gain of this communication-efficient filter. The optimal $\boldsymbol{M}_k^i$ is redesigned by minimizing the upper bound of the error covariance matrix $\boldsymbol{\bar P}_k^i$, following $\partial\text{tr}(\boldsymbol{\bar P}_k^i)/\partial\boldsymbol{M}_k^i=\boldsymbol{0}$. Specifically, the derivation of $\boldsymbol{\bar P}_k^i$ and $\boldsymbol{M}_k^i$  when $\gamma_k^i=0$ are presented in 
Appendix C, 
resulting in
\begin{equation}
    \begin{aligned}
       \boldsymbol{\bar P}_k^i=&\mu_1\boldsymbol{A}_k^i\boldsymbol{P}_{k\vert k-1}^i(\boldsymbol{A}_k^i)^{\rm T}+\mu_2\boldsymbol{M}_k^i\boldsymbol{R}_{k}^i(\boldsymbol{M}_k^i)^{\rm T}\\
       &+\mu_3\boldsymbol{M}_k^i\delta^i(\boldsymbol{M}_k^i)^{\rm T}, 
    \end{aligned}
\label{eq up_P 1}	
\end{equation}
\begin{equation*}
	\begin{aligned}
		\boldsymbol{M}_k^i=&\mu_1\boldsymbol{P}_{k\vert k-1}^i(\boldsymbol{{G}}_k^i)^{\rm T}\\
        &\times \left[ \mu_1\boldsymbol{{G}}_k^i\boldsymbol{P}_{k\vert k-1}^i(\boldsymbol{{G}}_k^i)^{\rm T}+\mu_2\boldsymbol{R}_k^i+\mu_3\delta^i\boldsymbol{I}\right]^{-1},
	\end{aligned}
	\label{eq gain 1}	
\end{equation*}
where $\mu_1,\mu_2,\mu_3$ are positive scalars following
\begin{equation}
    \mu_1=1+\sigma_1,\ \mu_2=1+\sigma_2,\  \mu_3=1+\sigma_1^{-1}+\sigma_2^{-1},
    \label{eq scaling parameter}
\end{equation}
and
\begin{equation*}
	\boldsymbol{A}_k^i=\boldsymbol{I}-\boldsymbol{M}_k^i\boldsymbol{{G}}_k^i,\ \boldsymbol{{G}}_k^i=\frac{\partial h^i(\boldsymbol{x}_k)}{\partial \boldsymbol{x}_k}\Big|_{\boldsymbol{x}_k=\boldsymbol{\hat x}_{k\vert k-1}^i}.
\end{equation*}
Therefore, the corresponding upper bound of the error covariance matrix is updated to
\begin{equation*}
	\boldsymbol{\bar P}_{k}^i=\gamma_k^i[\boldsymbol{P}_{k\vert k-1}^i-\boldsymbol{K}_k^i\boldsymbol{P}_{z_k,z_k}^i(\boldsymbol{K}_k^i)^{\rm T}]+(1-\gamma_k^i)\boldsymbol{\bar P}_k^i.
 \label{eq P upper bound}
\end{equation*}
\vspace{-0.4cm}
\begin{remark}
    When the transmission of $\boldsymbol{z}_k^i$ is triggered at instant $k$, i.e., $\gamma_k^i=1$, Eq.~\eqref{eq ET-CKF_upx} becomes $\boldsymbol{\hat x}_{k}^i=\boldsymbol{\hat x}_{k\vert k-1}^i+\boldsymbol{K}_k^i(\boldsymbol{z}_k^i-\boldsymbol{\hat z}_{k\vert k-1}^i)$, indicating that the ET filter turns into normal CKF. However, when the measurement transmission is not triggered, i.e., $\gamma_k^i=0$, Eq.~\eqref{eq ET-CKF_upx} changes to $\boldsymbol{\hat x}_{k}^i=\boldsymbol{\hat x}_{k\vert k-1}^i+\boldsymbol{M}_k^i(\boldsymbol{z}_{\tau^i_{k-1}}^i-\boldsymbol{\hat z}_{k\vert k-1}^i)$, where $\boldsymbol{M}_k^i$ is the filter gain for the non-triggered measurement update term.
\end{remark}
\begin{remark}
    Since $\boldsymbol{M}_k^i(\boldsymbol{M}_k^i)^{\rm T} \geq \boldsymbol{0}$, the upper bound of error covariance matrix $\boldsymbol{\bar P}_k^i$ given in \eqref{eq up_P 1} increases with the static trigger threshold $\delta^i$, indicating that the estimation error grows as the measurement transmission frequency is reduced. This theoretical analysis reveals an inevitable trade-off between the ET filter's tracking accuracy and communication rate, the performance of which is heavily influenced by the trigger threshold $\delta^i$.
\end{remark}

\subsubsection{Event-Triggered CIF}
For the convenience of further analysis, we define the pseudo measurement matrix $\boldsymbol{H}_k^i$ as
\begin{equation*}
	\boldsymbol{H}_k^i \triangleq (\boldsymbol{P}^i_{x_k,z_k})^{\rm T}(\boldsymbol{P}^i_{k\vert k-1})^{-1}=(\boldsymbol{P}^i_{x_k,z_k})^{\rm T}\boldsymbol{Y}^i_{k\vert k-1}.
	\label{eq H}
\end{equation*}
Then, the corresponding correlation information matrix $\boldsymbol{\bar I}_k^{i}$ and information contribution $\boldsymbol{\bar i}_k^{i}$ under the ET mechanism are computed by


\begin{equation}
	\left\{
	\begin{aligned}
		\boldsymbol{\bar I}_k^{i}=&\gamma_k^i(\boldsymbol{H}^i_{k})^{\rm T}(\boldsymbol{ D}_k^{i})^{-1}\boldsymbol{H}^i_{k}\\
		&-(1-\gamma_k^i)\boldsymbol{Y}^i_{k\vert k-1}[(\boldsymbol{N}_k^i)^{-1}+\boldsymbol{Y}^i_{k\vert k-1}]^{-1}\boldsymbol{Y}^i_{k\vert k-1},  \\
		\boldsymbol{\bar i}_k^{i}=&\gamma_k^i(\boldsymbol{H}^i_{k})^{\rm T}( \boldsymbol{D}_k^{i})^{-1}(\Delta\boldsymbol{z}_k^{i}\!+\boldsymbol{H}^i_{k}\boldsymbol{\hat x}_{k|k-1}^{i}) \\
		&+(1-\gamma_k^i)[\mu_1(\boldsymbol{G}_k^i)^{\rm T}(\boldsymbol{S}_k^i)^{-1}\Delta\boldsymbol{\bar z}_k^i\\
        &+\boldsymbol{\bar I}_k^{i}(\boldsymbol{\hat x}_{k\vert k-1}^i+\boldsymbol{M}_k^i\Delta\boldsymbol{\bar z}_k^i)],  
	\end{aligned}
	\right.
	\label{eq i and I with H}
\end{equation}
where $\Delta\boldsymbol{\bar z}_k^i=\boldsymbol{z}_{\tau^i_{k}}^i-\boldsymbol{\hat z}_{k\vert k-1}^i$ since each sensor's current measurement $\boldsymbol{z}_k^i$ is updated only when the specific triggering condition is met, and
\begin{equation*}
	\begin{aligned}
		\boldsymbol{S}_k^{i}&=\mu_1\boldsymbol{G}^i_k\boldsymbol{P}_{k\vert k-1}^i(\boldsymbol{G}^i_k)^{\rm T}+\mu_2\boldsymbol{R}_k^i+\mu_3\delta^i\boldsymbol{I},
	\end{aligned}
	\label{eq S}
\end{equation*}
\begin{equation*}
	\begin{aligned}
		\boldsymbol{N}_k^{i}=&(\sigma_1\boldsymbol{I}_{n\times n}-\mu_1\boldsymbol{M}^i_k\boldsymbol{G}^i_k)\boldsymbol{P}_{k\vert k-1}^i\\
        &-\mu_1\boldsymbol{P}_{k\vert k-1}^i(\boldsymbol{M}^i_k\boldsymbol{G}^i_k)^{\rm T}+\boldsymbol{M}^i_k\boldsymbol{S}^i_k(\boldsymbol{M}^i_k)^{\rm T}.
	\end{aligned}
	\label{eq N}
\end{equation*}
The detailed derivation of \eqref{eq i and I with H} is provided in 
Appendix D.
Consequently, the information vector $\boldsymbol{\hat y}_{k}^{i}$ and information matrix $\boldsymbol{Y}_{k}^{i}$ of ET-CIF are updated according to
\begin{equation}
	\boldsymbol{\hat y}_{k}^{i}=\boldsymbol{\hat y}_{k|k-1}^{i}+\boldsymbol{\bar i}_k^{i}, \ 
	\boldsymbol{Y}_{k}^{i}=\boldsymbol{Y}_{k|k-1}^{i}+\boldsymbol{\bar I}_k^{i}.
	\label{eq info_parir_ET}
\end{equation}
The ET-CIF algorithm at node $\nu_i$ is specified in Alg.~\ref{ET-CIF}. Particularly, the operation steps of ET-CIF without specific formula explanations are the same as those of the CKF algorithm in Appendix A.

\begin{remark}
    When $\gamma_k^i=1$, Eq.~\eqref{eq info_parir_ET} is equivalent to $\boldsymbol{\hat y}_{k}^{i}=\boldsymbol{\hat y}_{k|k-1}^{i}+\boldsymbol{i}_k^{i},\  \boldsymbol{Y}_{k}^{i}=\boldsymbol{Y}_{k|k-1}^{i}+\boldsymbol{I}_k^{i}$, indicating that ET-CIF becomes the normal CIF as given in Sec.~\ref{CIF}. For the measurement non-triggered cases where $\gamma_k^i=0$, the information contribution $\boldsymbol{\bar i}_k^{i}$ and correlation information matrix $\boldsymbol{\bar I}_k^{i}$ are designed according to the ET-CKF.
\end{remark}

\begin{remark}
    Note that the last transmitted measurement utilized in the non-triggered cases can be compensated by a random value following Gaussian distribution at each time instance, \textit{i.e}., $\boldsymbol{z}_{\tau_k^i}^i=\boldsymbol{z}_{\tau_{k-1}^i}^i+\boldsymbol{n}_k^i$, where $\boldsymbol{n}_k^i\sim\mathcal{N}(\boldsymbol{\mu}_i,\boldsymbol{\Sigma}_i^2)$, to reduce the accumulation of random biases caused by sensor noises. As the random value $\boldsymbol{n}_k^i$ is designed to approximate the measurement noise, the mean of its distribution $\boldsymbol{\mu}_i$ can be set as $\boldsymbol{0}$, and the variance value $\boldsymbol{\Sigma}_i^2$ is tuned according to the predefined trigger threshold $\delta^i$.
\end{remark}

\begin{remark}
    According to Lemma 2 in the supplementary material,
    the upper bound for the error covariance matrix $\boldsymbol{\bar P}_k^i$ of the designed ET-CIF algorithm is related to the scaling parameters $\mu_1,\mu_2,\mu_3$, thus may bring in conservativeness \cite{SECUKF}. 
    Such conservativeness can be alleviated by appropriately setting the values of $\sigma_1$ and $\sigma_2$, as given in~\eqref{eq scaling parameter}, which rationalizes the weights of modification terms in non-measurement-triggered cases. 
\end{remark}

\vspace{-0.32cm}
\begin{algorithm}
	\SetAlgoNoLine
	\caption{Event-Triggered CIF (ET-CIF)}\label{ET-CIF}
        \KwIn{initial target state $\boldsymbol{x}_0$, measurement $\boldsymbol{z}_k^i$.}
	\KwOut{information pair $(\boldsymbol{\hat y}_{k}^i, \boldsymbol{Y}_{k}^i)$ and estimation pair $(\boldsymbol{\hat x}_{k}^i, \boldsymbol{P}_{k}^i)$ at each time instant $k$.}
	\textbf{Initialization:} $\boldsymbol{\hat x}_{0}^i=\mathbb{E}\{\boldsymbol{x}_0\},\  \boldsymbol{P}_0^i=\mathbb{E}\{(\boldsymbol{x}_0-\boldsymbol{\hat x}_{0}^i)(\boldsymbol{x}_0-\boldsymbol{\hat x}_{0}^i)^{\rm{T}}\}.$ \\
	\For{$k\leftarrow 1$ \rm{to} $T$}
	{\textbf{Phase 1: Prediction} \\
		$\boldsymbol{x}_{k-1}^{i,r},\  r=1,\dots,2n\ \leftarrow\ $ Generate $2n$ points.\\
		$\boldsymbol{\hat x}_{k\vert k-1}^i,\ \boldsymbol{P}_{k\vert k-1}^i\ \leftarrow\ $ Calculate prior values. \\
		\textbf{Phase 2: Update} \\
		$\boldsymbol{\hat z}_{k\vert k-1}^i\ \leftarrow\ $ Obtain the predicted measurement. \\
		$\boldsymbol{P}_{z_k,z_k}^i,\  \boldsymbol{P}_{x_k,z_k}^i\ \leftarrow\ $ Compute covariance matrices.   \\
		$\gamma_k^i\ \leftarrow\ $ Determine the trigger flag from \eqref{eq gamma}.\\
		$\boldsymbol{\bar i}_k^{i},\  \boldsymbol{\bar I}_k^{i}\ \leftarrow\ $Compute the information contribution and correlation information matrix by \eqref{eq i and I with H}. \\
		$\boldsymbol{\hat y}_{k}^{i},\  \boldsymbol{Y}_{k}^{i}\ \leftarrow\ $ Calculate the posterior information vector and matrix based on \eqref{eq info_parir_ET}. \\
		$\boldsymbol{\hat x}_{k}^{i},\  \boldsymbol{P}_{k}^i\ \leftarrow\ $ Convert to the posterior estimate and error covariance matrix according to \eqref{eq posterior_xp}.
	}
\end{algorithm}
\vspace{-0.3cm}

\vspace{-0.35cm}
\subsection{Diffusion-based Filter} \label{ED-CIF}
Based on the single-sensor nonlinear filter, a diffusion-based collaborative filter obtains more accurate and robust global estimation results by fusing the information contribution and posterior estimate of each node and its neighboring nodes. In the adaptation stage of diffusion-based CIF, each sensor node's information pair is modified by
\begin{equation}
	\left\{
	\begin{aligned}
		\boldsymbol{\hat y}_{k}^{i}&=\boldsymbol{\hat y}_{k|k-1}^{i}+\boldsymbol{\bar i}_k^{i}+\sum_{\nu_j\in\boldsymbol{\mathcal{N}}_i}\boldsymbol{\bar i}_k^{j}, \\
		\boldsymbol{Y}_{k}^{i}&=\boldsymbol{Y}_{k|k-1}^{i}+\boldsymbol{\bar I}_k^{i}+\sum_{\nu_j\in\boldsymbol{\mathcal{N}}_i}\boldsymbol{\bar I}_k^{j},
	\end{aligned}
	\right.
	\label{eq diff_ada}
\end{equation}
from which, the posterior estimate $\boldsymbol{\hat x}_{k}^{i}$ and error covariance matrix $\boldsymbol{P}_{k}^{i}$ are obtained from \eqref{eq posterior_xp}. 
As for the combination stage, the local posterior estimate and covariance matrix at each sensor node are fused by
\begin{equation}
	\left\{
	\begin{aligned}
		\boldsymbol{\hat x}_{k,\text{fus}}^{i}&=c_k^{i,i}\boldsymbol{\hat x}_{k}^{i}+\sum_{\nu_j\in\boldsymbol{\mathcal{N}}_i}c_k^{i,j}\boldsymbol{\hat x}_{k}^{j}, \\
        \boldsymbol{P}_{k,\text{fus}}^{i}&=c_k^{i,i}\boldsymbol{P}_{k}^{i}+\sum_{\nu_j\in\boldsymbol{\mathcal{N}}_i}c_k^{i,j}\boldsymbol{P}_{k}^{j},
	\end{aligned}
	\right.
	\label{eq diff_com}
\end{equation}
where $\boldsymbol{C}_k=[c_k^{i,j}]_{N\times N}$ is the row-stochastic diffusion weight matrix following
\begin{equation*}
	\boldsymbol{C}_k\boldsymbol{1}_{N\times 1}=\boldsymbol{1}_{N\times 1}, 
	\label{eq C}
\end{equation*}
where $c_k^{i,j}=0 \ \text{if}\ \nu_i\notin\boldsymbol{\mathcal{N}}_j \cap \nu_i\neq\nu_j$, $c_k^{i,j}\geq0 \ \text{for}\ \forall \nu_i,\nu_j$, and $\boldsymbol{1}_{N\times 1}$ is an $N\times 1$ column vector with unity entries. Specifically, the calculation rule of $c_k^{i,j}$ follows
\begin{equation} \label{eq wk}
	c_k^{i,j}=\left\{ 
	\begin{array}{l}
		\frac{1}{1+\Delta},\ \ \text{if}\  \nu_j\in\boldsymbol{\mathcal{N}}_i, \\
		1-\sum\limits_{\nu_j\in\boldsymbol{\mathcal{N}_i}}c_k^{i,j},\ \ \text{if}\  \nu_i=\nu_j,\\
		0,\ \ \text{otherwise},
	\end{array}
	\right. 
\end{equation}
where $\Delta=\text{max}_{\nu_i\in\boldsymbol{\mathcal{V}}}|\mathcal{N}_i|$ denotes the maximum degree of all the sensor nodes. 

\begin{remark}
    For diffusion-based CIF, the transmitted data in the adaptation stage is no longer the measurement but the information contribution and correlation information matrix that contains the updated observation messages to avoid
the transfer of high-dimensional measurement matrices. 
In the combination stage, the transmitted data is converted back to the estimate and error covariance matrix to obtain the final fused results.
\end{remark}


\begin{remark}
   Unlike conventional consensus-based fusion schemes that require a large number of iterations to reach an agreement among sensor nodes, the fusion operation in diffusion strategy only consists of one iteration as in \eqref{eq diff_ada} and \eqref{eq diff_com}. Assume the adequate consensus iteration number as $L$. Then, for node $\nu_i$ with $|\mathcal{N}_i|$ adjacent nodes, 
the amounts of data for calculating the target state are $nL|N_i|$ and $2n|N_i|$, respectively, for consensus and diffusion schemes, indicating that the data transmission is reduced in diffusion schemes since $2n<nL$ in most cases.
\end{remark}

\vspace{-0.35cm}
\subsection{Covariance Intersection Fusion} \label{CIF-CI}
The estimate fusion in \eqref{eq diff_com} is given as a convex combination form without taking cross-covariance into consideration, which may deteriorate the filter's performance since the cross-correlations between local estimates from different nodes are typically unknown. Therefore, the covariance intersection (CI) fusion \cite{CI} is considered in the diffusion-based collaborative filter to further improve the global estimate's accuracy with interacting combination weights.
In this case, the inverse of the fused error covariance matrix for each node $\nu_i\in\boldsymbol{\mathcal{V}}$ is computed in advance according to
\begin{equation}
	(\boldsymbol{P}_{k,\text{fus}}^{i})^{-1}=c_k^{i,i}(\boldsymbol{P}_{k}^{i})^{-1}+\sum_{\nu_j\in\boldsymbol{\mathcal{N}}_i}c_k^{i,j}(\boldsymbol{P}_{k}^{j})^{-1}.
	\label{eq diff_com_Y_ci}
\end{equation}
Then, the fused estimate in \eqref{eq diff_com} is updated to
\begin{equation}
	\boldsymbol{\hat x}_{k,\text{fus}}^{i}=\boldsymbol{P}_{k,\text{fus}}^i[c_k^{i,i}(\boldsymbol{P}_{k}^i)^{-1}\boldsymbol{\hat x}_{k}^{i}+\sum_{\nu_j\in\boldsymbol{\mathcal{N}}_i}c_k^{i,j}(\boldsymbol{P}_{k}^j)^{-1}\boldsymbol{\hat x}_{k}^{j}].
	\label{eq diff_com_x_ci}
\end{equation}
The event-triggered diffusion-based cubature information filter with covariance intersection, abbreviated as EDC-CIF\footnote{The stability analysis of EDC-CIF is presented in Appendix E through a proof of the mean-square stochastic boundedness of the estimation errors.}, is elaborated in Alg.~\ref{ED-CIF-CI}. 

\begin{remark}
    As calculated in \eqref{eq diff_com_x_ci}, the convex combination of posterior estimates is weighted according to the error covariance matrix to obtain the final fused result. The weight $(\boldsymbol{P}_{k}^j)^{-1}$ utilized here indicates that the corresponding estimate with a smaller error covariance matrix counts more in the data fusion process, thus enhancing the distributed fusion's estimation accuracy and reliability.
\end{remark}

\begin{remark}
    In the perspective of information filter, Eq.~\eqref{eq diff_com_Y_ci} and Eq.~\eqref{eq diff_com_x_ci} are equivalent to $\boldsymbol{Y}^i_{k,\text{fus}}=c_k^{i,i}\boldsymbol{Y}^i_k+\sum_{\nu_j\in\mathcal{N}_i}c_k^{i,j}\boldsymbol{Y}^j_k$ and $\boldsymbol{\hat y}^i_{k,\text{fus}}=c_k^{i,i}\boldsymbol{\hat y}^i_k+\sum_{\nu_j\in\mathcal{N}_i}c_k^{i,j}\boldsymbol{\hat y}^j_k$, respectively. This illustrates that the combination operation for the estimation pair $(\boldsymbol{\hat x}_k^i, \boldsymbol{P}_k^i)$ in the diffusion strategy with CI is the same as the direct combination of the information pair $(\boldsymbol{\hat y}_k^i, \boldsymbol{Y}_k^i)$.
\end{remark}

\begin{remark}
    The computational overhead of EDC-CIF is characterized by a lightweight ET check of $\mathcal{O}(m)$ and a non-iterative Diffusion fusion step of $\mathcal{O}(n^2 |\mathcal{N}_i|)$. Unlike consensus-based methods that require $L > 1$ iterations, the proposed scheme eliminates the $L$-fold complexity factor, ensuring high efficiency. The marginal investment in computation enables a substantial reduction in communication, validating the favorable trade-off between local processing and network-wide information sharing.
\end{remark}


\begin{algorithm}[htbp]
	\SetAlgoNoLine
	\caption{Event-Triggered Diffusion-based CIF with Covariance Intersection (EDC-CIF)}\label{ED-CIF-CI}
        \KwIn{initial target state $\boldsymbol{x}_0$, measurement $\boldsymbol{z}_k^i$.}
	\KwOut{fused information pair $(\boldsymbol{\hat y}_{k,\text{fus}}^i, \boldsymbol{Y}_{k,\text{fus}}^i)$ and fused estimation pair $(\boldsymbol{\hat x}_{k,\text{fus}}^i, \boldsymbol{P}_{k,\text{fus}}^i)$ at each time instant $k$.}
	\textbf{Initialization:} $\boldsymbol{\hat x}_0^i=\mathbb{E}\{\boldsymbol{x}_0\},\ \boldsymbol{P}_0^i=\mathbb{E}\{(\boldsymbol{x}_0-\boldsymbol{\hat x}_0^i)(\boldsymbol{x}_0-\boldsymbol{\hat x}_0^i)^{\rm{T}}\}.$ \\
	\For{$k\leftarrow 1$ \rm to $T$}
	{\textbf{Local Phase: Event-triggered CIF} \\
		$\boldsymbol{\bar i}_k^{i},\ \boldsymbol{\bar I}_k^{i}\ \leftarrow\ $ Compute the information contribution and correlation information matrix as in Alg.~\ref{ET-CIF}. \\
		$\boldsymbol{\hat y}_{k}^{i},\ \boldsymbol{Y}_{k}^{i}\ \leftarrow\ $ Calculate the posterior information vector and matrix according to \eqref{eq info_parir_ET}. \\
		\textbf{Global Phase: Diffusion fusion with CI} \\
		\textbf{Stage 1: Adaptation} \\
		$\boldsymbol{\hat y}_{k}^{i},\ \boldsymbol{Y}_{k}^{i}\ \leftarrow\ $ Modify the posterior information vector and information matrix based on \eqref{eq diff_ada}. \\
		$\boldsymbol{\hat x}_{k}^{i},\ \boldsymbol{P}_{k}^i\ \leftarrow\ $ Convert to the posterior estimate and error covariance matrix following \eqref{eq posterior_xp}. \\
		\textbf{Stage 2: Combination} \\
		$\boldsymbol{P}_{k,\text{fus}}^i\ \leftarrow\ $ Obtain the fused error covariance matrix by \eqref{eq diff_com_Y_ci}, and set $\boldsymbol{P}_{k}^i=\boldsymbol{P}_{k,\text{fus}}^i$. \\
		$\boldsymbol{\hat x}_{k,\text{fus}}^{i}\ \leftarrow\ $ Determine the fused estimate by \eqref{eq diff_com_x_ci}, and set $\boldsymbol{\hat x}_{k}^{i}=\boldsymbol{\hat x}_{k,\text{fus}}^{i}$. \\
        $\boldsymbol{\hat y}_{k,\text{fus}}^i,\ \boldsymbol{Y}_{k,\text{fus}}^i \ \leftarrow\ $ Convert to the corresponding fused information vector and matrix from
        $\boldsymbol{\hat y}_{k,\text{fus}}^i=(\boldsymbol{P}_{k,\text{fus}}^i)^{-1}\boldsymbol{\hat x}_{k,\text{fus}}^i,\ \boldsymbol{Y}_{k,\text{fus}}^i=(\boldsymbol{P}_{k,\text{fus}}^i)^{-1}.$ \\
        
	}
\end{algorithm}

\section{Evaluation} \label{evaluation}

{This section conducts both numerical simulations and real-world experiments with various types of sensors, such as LiDAR and RGB-D camera, to evaluate the generalized tracking performance of Alg.~\ref{ED-CIF-CI}.} 
Our method is compared with a series of {advanced ET collaborative nonlinear filters} to show its superiority in estimation accuracy, convergence speed, and communication efficiency.
{We also conduct ablation studies to validate the effectiveness of the proposed modules} and analyze the key factors influencing the algorithm's performance.

\vspace{-0.35cm}
\subsection{Comparative Methods}
The following baselines are considered in the simulation. 
\begin{itemize}
    \item Consensus-based CKF (\textbf{C-CKF}) \cite{CCKF}: a conventional distributed nonlinear filter that does not conserve communication resources or degrade tracking performance. 
    \item ET consensus-based EKF/ UKF/ CKF (\textbf{EC-EKF/ EC-UKF/ EC-CKF}) \cite{SECEKF, SECUKF, SECCKF}: existing distributed nonlinear filters with low data transmission.
    \item Diffusion-based CIF (\textbf{D-CIF}) \cite{wang2018diffusion}: an existing diffusion-based distributed nonlinear filter without communication resource saving. 
    \item Diffusion-based CIF with CI (\textbf{DC-CIF}): our developed D-CIF considering unknown sensor correlations.
    \item ET diffusion-based CIF (\textbf{ED-CIF}): our developed communication-efficient distributed nonlinear filter. 
\end{itemize}
Note that our proposed algorithm, \textbf{EDC-CIF}, is the abbreviation of ED-CIF with CI. Complete performance comparison results can be found in Appendix F-I.

\subsection{Simulation Settings}
Take the state vector at the $k$th time instant as 
\begin{equation*}
    \boldsymbol{x}_k=\left[x_k,\ y_k,\ z_k,\ \dot x_k,\ \dot y_k,\ \dot z_k,\ \omega_k\right]^{\rm{T}},
\end{equation*}
where $\left(x_k,\ y_k,\ z_k \right) $ represents the target position in three-dimensional space, $\left(\dot x_k,\ \dot y_k,\ \dot z_k\right) $ denotes the three-direction velocity, and $\omega_k$ is the unknown constant turn rate. In the simulation, the target trajectory is modeled as an ascending circular motion. 
The measurement vector is given as $\boldsymbol{z}_k^i=[r_k^i,\ \phi_k^i,\ \rho_k^i]^{\rm{T}} $, where $r_k^i$ represents the relative distance between the target and sensor $\nu_i$, $\phi_k^i$ is the pitch angle, and $\rho_k^i$ denotes the azimuth angle. Given the location of each sensor $\nu_i$ as $\left(x_s^i,\ y_s^i,\ z_s^i \right)$, the observation function follows
\begin{equation}
	\left\{ 
	\begin{array}{l}
		r_k^i=\sqrt{(x_k-x_s^i)^2+(y_k-y_s^i)^2+(z_k-z_s^i)^2} + v_r^i, \\
		\phi_k^i=\arctan\left( \frac{z_k-z_s^i}{\sqrt{(x_k-x_s^i)^2+(y_k-y_s^i)^2}}\right) + v_{\phi}^i, \\
		\rho_k=\arctan\left( \frac{y_k-y_s^i}{x_k-x_s^i} \right) + v_{\rho}^i,
	\end{array}
	\right. \label{eq measure}
\end{equation}
where $(v_r^i,\ v_{\phi}^i,\ v_{\rho}^i)$ stands for the corresponding additive white Gaussian noises of the three measurement variables for each sensor $\nu_i$. 
{Note that \eqref{eq measure} corresponds to the measurement model of the LiDAR sensor mounted on the UAV in our simulation scenario.}
Specifically, we represent the sensor network as a connected undirected graph with $4$ nodes,
i.e., $\mathcal{V}=\left\lbrace\nu_1,\ \nu_2,\ \nu_3,\ \nu_4 \right\rbrace$. 
Based on the given communication topologies, the diffusion weight matrix $\boldsymbol{C}_k$ is calculated according to \eqref{eq wk}. 

{\subsection{Real-world Experimental Setup}} 
\textcolor{black}{In the real-world experiments, we adopt a configuration consisting of three observer UAVs equipped with RGB-D cameras and one target UAV, which provides a representative yet controllable cooperative sensing scenario. The three-observer system captures key characteristics of cooperative multi-UAV perception, including multi-view sensing, distributed information fusion, and cross-perspective consistency \cite{olfati2007distributed}. In particular, three spatially distributed observers form a minimal configuration enabling reliable triangulation and cooperative state estimation under sensing uncertainties \cite{zheng2025optimal}.\footnote{\textcolor{black}{From an experimental perspective, this setup also provides a practical balance between system complexity and operational safety. Coordinated multi-UAV flights introduce challenges in communication reliability, synchronization, and collision avoidance. Therefore, following common practice in multi-UAV research \cite{zheng2025optimal, zhou2022integrated, gu2018multiple, stasinchuk2021multi, zhu2020multi, li2022resilient}, real-world experiments are used to validate system feasibility, while scalability is further examined through larger-scale simulations \cite{tian2024ucdnet, hu2022where2comm, zheng2025optimal}.}}.
}
\subsubsection{Implementation Details}
{The target UAV used in our experiments is a quadrotor DJI Mini 3 Pro, serving as a moving object for 3D localization. Each of the three observer UAVs is equipped with an Intel D435i depth camera and an onboard Intel Core i7 NUC for processing depth data and computing the target's location, as shown in Fig.~\ref{devices}. Ground truth is obtained using a motion capture system covering an $8,\text{m} \times 8,\text{m} \times 2.5,\text{m}$ space, which records the 3D positions and quaternions of all UAVs. Each observer localizes the UAV target by detecting its 2D image position from the D435i's RGB stream and computing its 3D coordinates. Timestamps from the three UAVs and the motion capture system are independently recorded and synchronized during preprocessing.}

\begin{figure}[htbp]
	\centering
	\includegraphics[width = 0.85\linewidth]{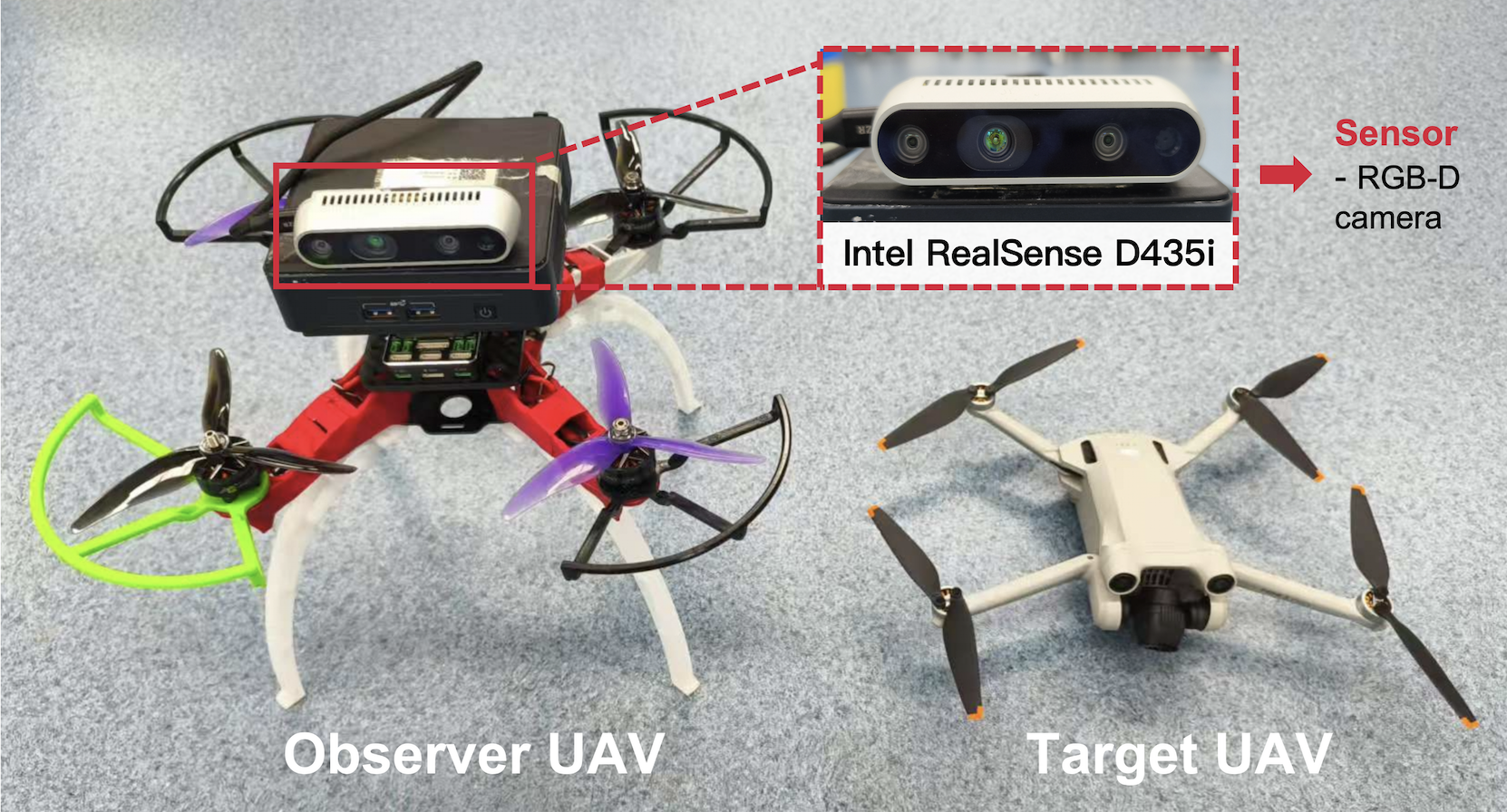}
	\caption{Implementation Devices. The left drone is one of the three observer UAVs, each equipped with an Intel RealSense D435i RGB-D camera for target recognition and distance measurement. The right drone is the dynamic target UAV, a DJI Mini 3 Pro, exhibiting nonlinear motion.}
	\label{devices}
\end{figure}

\subsubsection{Drone Trajectories}
{Two distinct target motion trajectories were evaluated to assess the robustness and adaptability of the proposed algorithm. The first trajectory followed an irregular, stochastic path generated via manual control, representative of real-world UAV operations. This included initial ascent, slight descent, and quasi-rectilinear motion across rectangular zones. The second trajectory was a programmatically controlled compound circular path (average radius: 0.7 m), consisting of concentric loops. These varied motion patterns help to ensure that our evaluation covers both unpredictable and structured flight behaviors.}

\subsubsection{Data Pre-processing}
\textcolor{black}{
Before filtering, we construct a comprehensive data preprocessing pipeline to ensure the reliability, consistency, and usability of the collected data. First, all raw trajectories, encompassing both measurements and ground truth values obtained from the motion capture system, are converted into the TUM format \cite{sturm2012benchmark} to ensure standardized data representation for subsequent processing. An outlier removal module is then applied to eliminate samples with abnormal temporal intervals and origin outliers, thereby improving data quality under practical communication conditions. 
To resolve clock discrepancies and asynchronous sampling across devices, timestamp alignment is performed by identifying consistent start and end times among all agents. Spatial trajectory alignment is subsequently carried out using the EVO evaluation tool \cite{grupp2017evo} to ensure consistent coordinate frames. Afterward, the ground-truth trajectory is downsampled to the same nominal rate as the observation data. Finally, interpolation-based frame-rate alignment is applied to synchronize the temporal resolution between the two trajectory streams.
%
Specifically, when the frame-rate discrepancy between the transmitted and received signals is mild, packet loss can be effectively mitigated through interpolation and temporal alignment. Otherwise, the network connection is checked.
By mitigating the transmission errors, spurious measurements, and unstable sampling such as delays and jitter, this preprocessing pipeline establishes reliable temporal–spatial synchronization signals for downstream filtering and tracking. 
}
\color{black}
\subsection{Nonlinear Tracking}
The ground truth and estimated trajectories of EDC-CIF are shown in Fig.\ref{track}, demonstrating accurate and stable tracking over time. Fig.\ref{xyzv} further validates the estimation performance by comparing the true and estimated 3D target states, including position and velocity, which converge with diminishing error. These results confirm the effectiveness of the diffusion-based CI fusion strategy for distributed nonlinear target tracking.



\begin{figure}[htbp]
	\centering
	\includegraphics[scale=0.35]{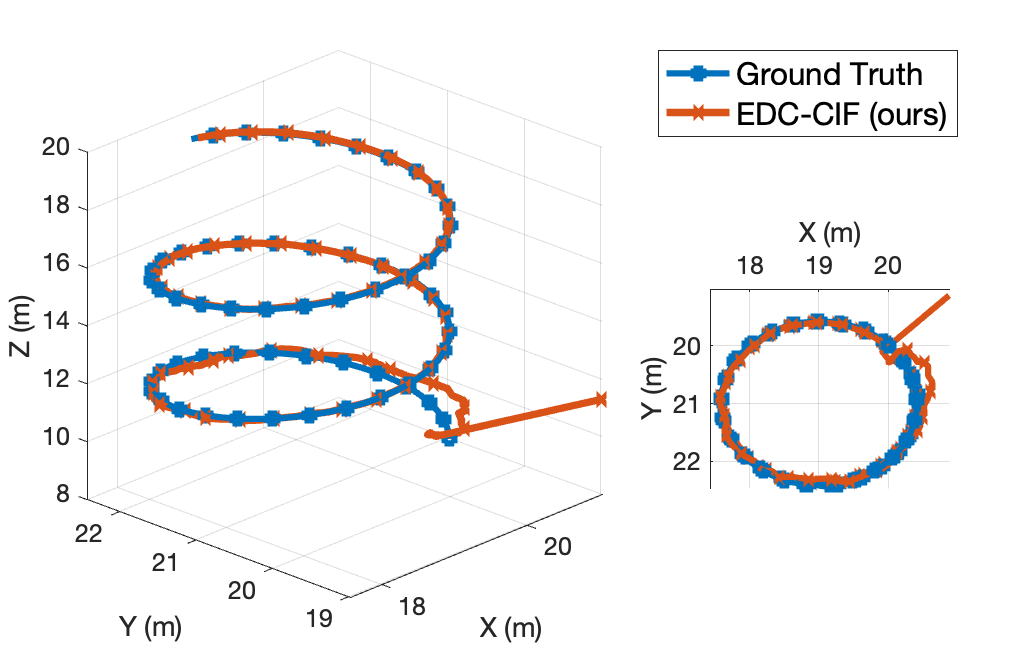}
    \caption{Trajectory comparison of the ground truth and EDC-CIF (numerical simulation).}
	\label{track}
\end{figure}

\begin{figure}[htbp]
    \centering
    \includegraphics[width=9cm, height=4.5cm]{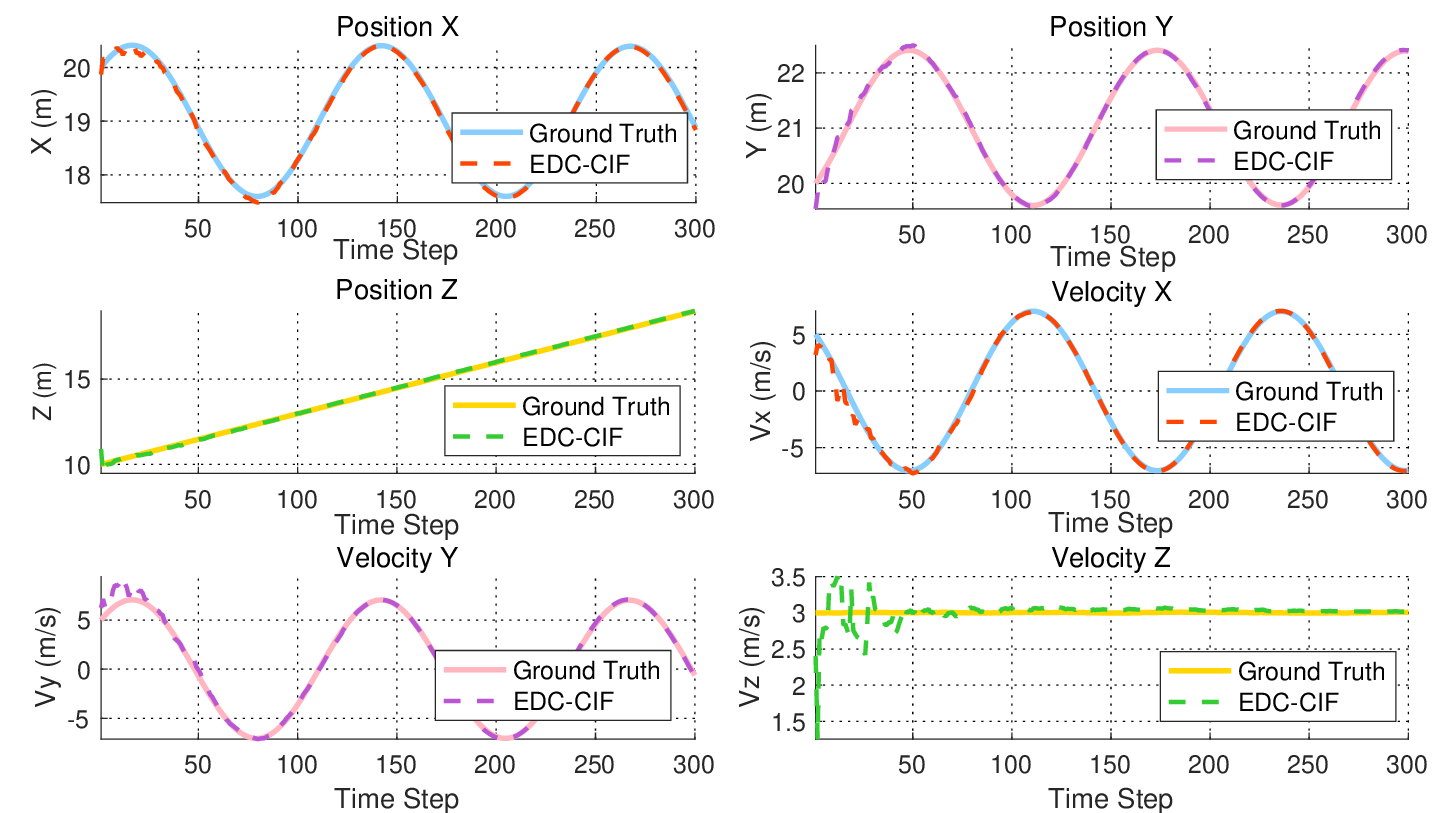}
    \caption{Ground truth and estimated state of EDC-CIF (numerical simulation).}
    \label{xyzv}
    \vspace{-0.3cm}
\end{figure}

{For real-world evaluation, the manually and programmatically controlled UAV target trajectories are shown in Fig.~\ref{track_manual} and Fig.~\ref{track_program}, respectively. The tracking results validate the effectiveness and robustness of our algorithm in real-world collaborative target tracking, as it closely approximates the ground truth despite noisy measurements. Comparison between the true target state and the EDC-CIF estimates in Fig.~\ref{xyz_manual} and Fig.~\ref{xyz_program} shows that EDC-CIF accurately estimates the target’s state, even for complex 3D movements. These results demonstrate that EDC-CIF generalizes well and adapts effectively to real-world collaborative UAV tracking scenarios.}
\begin{figure}[htbp]
\centering
\subfigure[Trajectory comparison]
{ \label{track_manual}
\includegraphics[width=0.47\columnwidth]{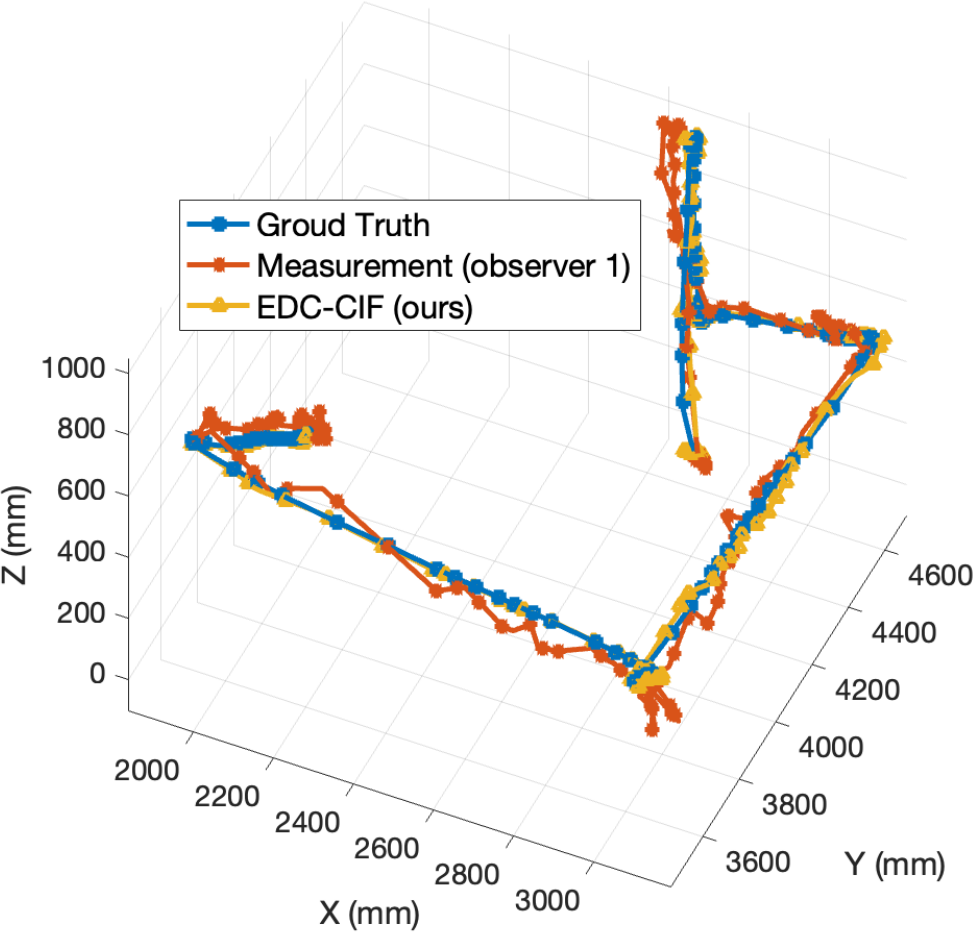}
}
\subfigure[State variable comparison]
{ \label{xyz_manual}
\includegraphics[width=0.46\columnwidth]{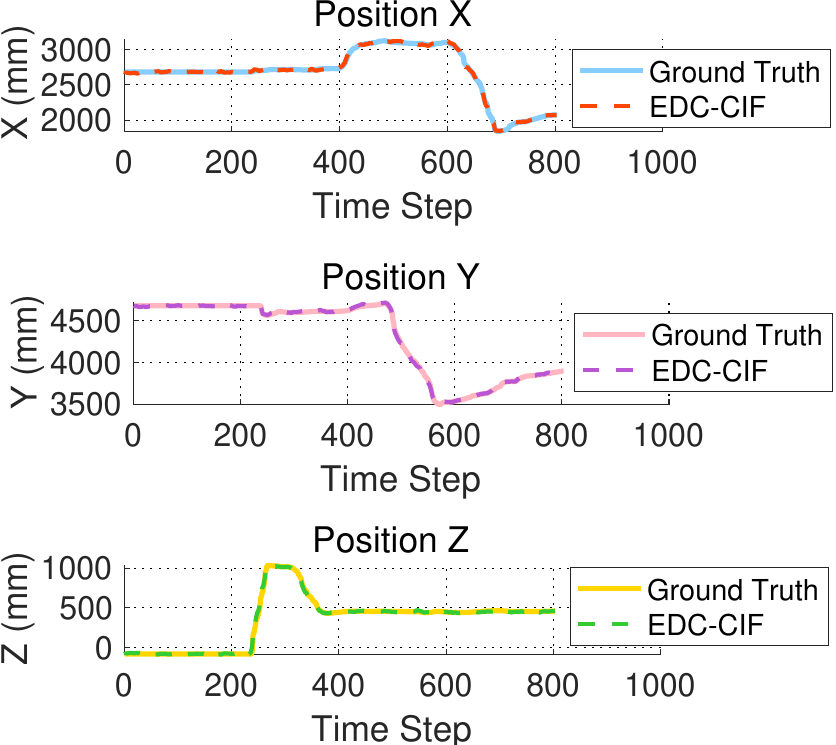}
}
\caption{Comparison of the ground truth and estimates of EDC-CIF (real-world manually controlled UAV experiment).}
\label{com_manual}
\vspace{-0.4cm}
\end{figure}

\begin{figure}[htbp]
\centering
\subfigure[Trajectory comparison]
{ \label{track_program}
\includegraphics[width=0.4\columnwidth]{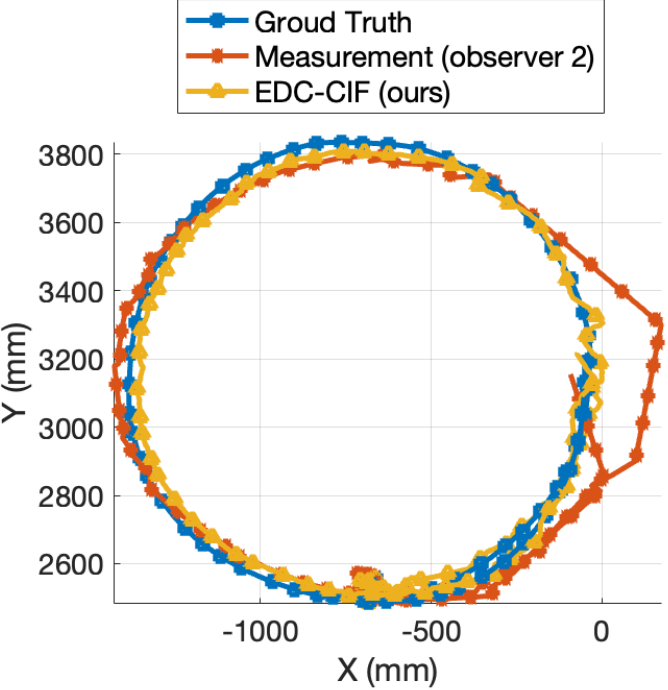}
}
\subfigure[State variable comparison]
{ \label{xyz_program}
\includegraphics[width=0.47\columnwidth]{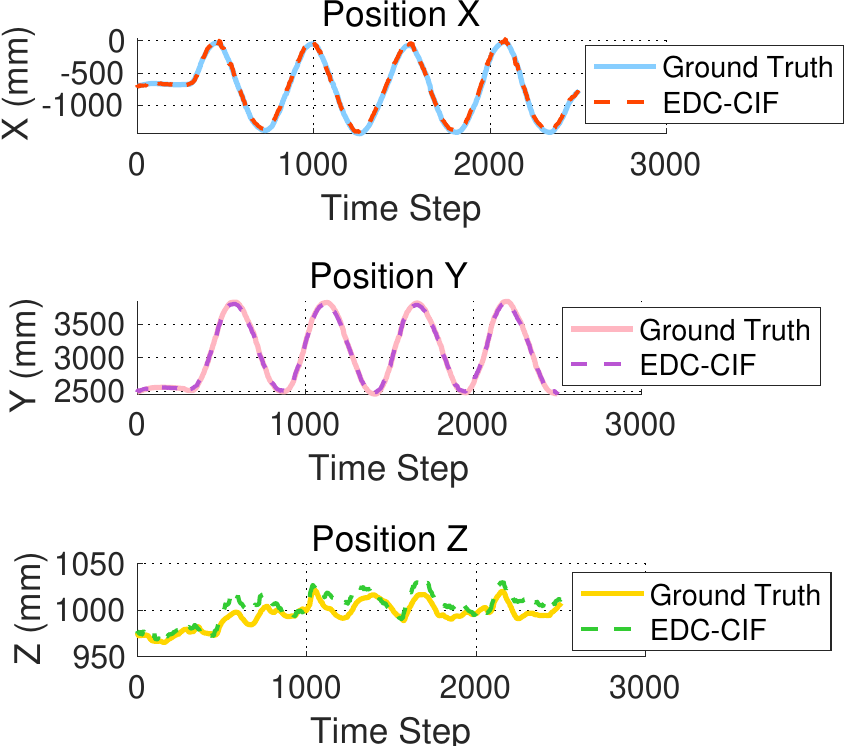}
}
\caption{Comparison of the ground truth and estimates of EDC-CIF (real-world programatically controlled UAV experiment).}
\label{com_program}
\vspace{-0.4cm}
\end{figure}

\subsection{Estimation Accuracy}
To quantify the estimation accuracy, the position root mean squared errors (RMSEs) and velocity RMSEs of conventional EC-CKF and EDC-CIF are compared in Fig.~\ref{RMSE}. Specifically, the position and velocity RMSE follow
\begin{equation*}
    \text{RMSE}_k^\text{pos}=\frac{1}{N}\sum\limits_{i=1}^{N}e_k^{i,\text{pos}}, \ \text{RMSE}_k^\text{vel}=\frac{1}{N}\sum\limits_{i=1}^{N}e_k^{i,\text{vel}},
    \label{eq RMSE}
\end{equation*}
where
\small
\begin{equation*}\label{eq PRMSE}
	e_k^{i,\text{pos}}\!\!\!=\!\!\!\sqrt{\!\frac{1}{MC}\!\!\sum\limits_{m=1}^{MC}\!\!\left[  \left( x_k^{i,m}\!-\!\hat{x}_k^{i,m}\right)^2\!\!\!\!\!+\!\left( y_k^{i,m}\!-\!\!\hat{y}_k^{i,m}\right)^2\!\!\!\!
		+\!\left(z_k^{i,m}\!-\!\hat{z}_k^{i,m}\right)^2\!\right]},
\end{equation*}
\begin{equation*}\label{eq VRMSE}
	e_k^{i,\text{vel}}\!\!\!=\!\!\!\sqrt{\!\frac{1}{MC}\!\!\sum\limits_{m=1}^{MC}\!\left[\!\left( \dot{x}_k^{i,m}\!-\!\hat{\dot{x}}_k^{i,m}\right)^2\!\!\!\!\!+\!\!\left( \dot{y}_k^{i,m}\!\!-\!\!\hat{\dot{y}}_k^{i,m}\right)^2\!\!\!
		\!\!+\!\!\left(\dot{z}_k^{i,m}\!\!-\!\hat{\dot{z}}_k^{i,m}\right)^2\!\right]},
\end{equation*}
\\
and $MC$ is the total number of Monte Carlo runs. 
It can be observed that EDC-CIF performs better than EC-CKF with smaller estimation errors and faster convergence in terms of both position and velocity variables, especially for the velocity ones. This demonstrates that the diffusion strategy outperforms the consensus scheme in tracking accuracy, indicating our method's efficiency in compensating for the accuracy loss induced by the ET mechanism.
\begin{figure} 
\centering
\subfigure[Position RMSE]
{ \label{fig:a}
\includegraphics[width=0.46\columnwidth]{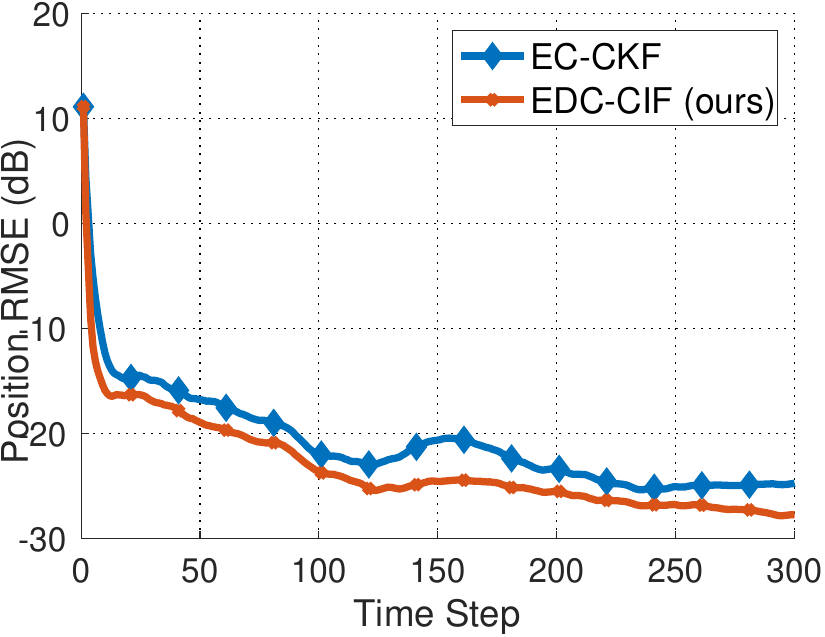}
}
\subfigure[Velocity RMSE]
{ \label{fig:b}
\includegraphics[width=0.46\columnwidth]{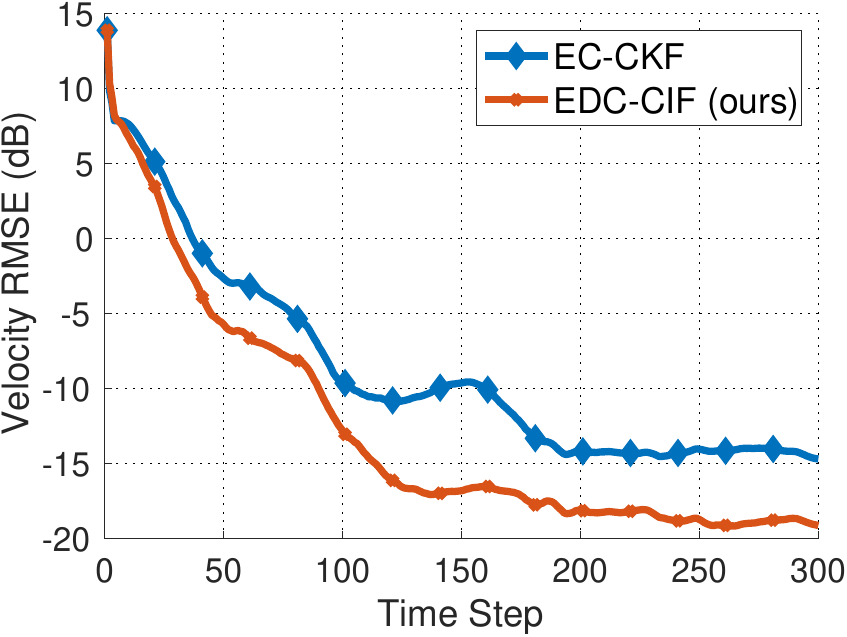}
}
\caption{Comparison of position and velocity RMSE.}
\label{RMSE}
\vspace{-0.4cm}
\end{figure}





Table~\ref{tracking_comparison_4} compares the means of RMSE for each state variable over 200 time instants among EC-EKF, EC-UKF, EC-CKF, ED-CIF, and EDC-CIF. 
The estimation error results demonstrate that the proposed EDC-CIF algorithm achieves the highest target tracking accuracy among existing state-of-the-art nonlinear communication-efficient collaborative filters, yielding the lowest RMSE for each target state variable. {Notably, it achieves up to a 66.49\% reduction in position estimation error}, highlighting its superiority in estimation accuracy and effectiveness in balancing reduced data transmission with global tracking performance. {More RMSE results for the real-world UAV experiments can be found in Appendix G in the supplementary material.}

\setlength{\tabcolsep}{2mm}
\begin{table}[htbp]
    \begin{threeparttable}[b]
	\caption{Average RMSE for each state variable.}
	\begin{tabular}{cccccc}
		\toprule
		State & EC-EKF  & EC-UKF & EC-CKF & ED-CIF & EDC-CIF\\
		\midrule
		$x$ (m)  & $0.2734$ & $0.0873$ & $0.0886$ & $0.0647$ & $\mathbf{0.0645}$
        \\  [2pt]
		$y$ (m)  & $0.2225$ & $0.0863$ & $0.0828$ & $0.0618$ & $\mathbf{0.0617}$ \\  [2pt]
        $z$ (m)  & $0.1331$ & $0.0857$ & $0.0809$ & $0.0656$ & $\mathbf{0.0655}$ \\  [2pt]
        $\dot x$ (m/s)  & $1.4747$ & $0.5343$ & $0.4936$ & $0.3704$ & $\mathbf{0.3685}$ \\  [2pt]
        $\dot y$ (m/s)  & $1.4103$ & $0.4158$ & $0.3976$ & $0.3021$ & $\mathbf{0.3007}$ \\  [2pt]
        $\dot z$ (m/s)  & $0.3212$ & $0.2571$ & $0.2365$ & $0.2075$ & $\mathbf{0.2067}$ \\  [2pt]
        $\omega$ (rad/s)  & $0.5816$ & $0.1961$ & $0.1875$ & $0.1617$ & $\mathbf{0.1613}$ \\  [2pt]
		\bottomrule
	\end{tabular}
 \label{tracking_comparison_4}
 \end{threeparttable}
 \vspace{-2em} 
\end{table} 

\subsection{Computation Time}
The total computation time $t_{\text{com}}$ of six distributed filters based on consensus or diffusion strategies during the whole simulation with 4 and 8 sensor nodes are depicted in Fig.~\ref{computation time}. It can be seen that all diffusion-based schemes take less running time than consensus-based methods, indicating their fast convergence speed and small computation consumption. It should be noted that, when sensor node number $N$ is $4$, there exists a $47.20\%$ reduction in the computation time of EDC-CIF compared to conventional EC-CKF, while the reduction percentage of time consumption increases to nearly $62.55\%$ when the node number equals $8$.
This reveals the great potential of diffusion-based collaborative filters in real-time collaborative state estimation assignments, especially for the ones with large-scale networks.

\begin{figure}[htbp]
	\centering
	\subfigure[Sensor node number $N=4$]{\includegraphics[scale=0.4]{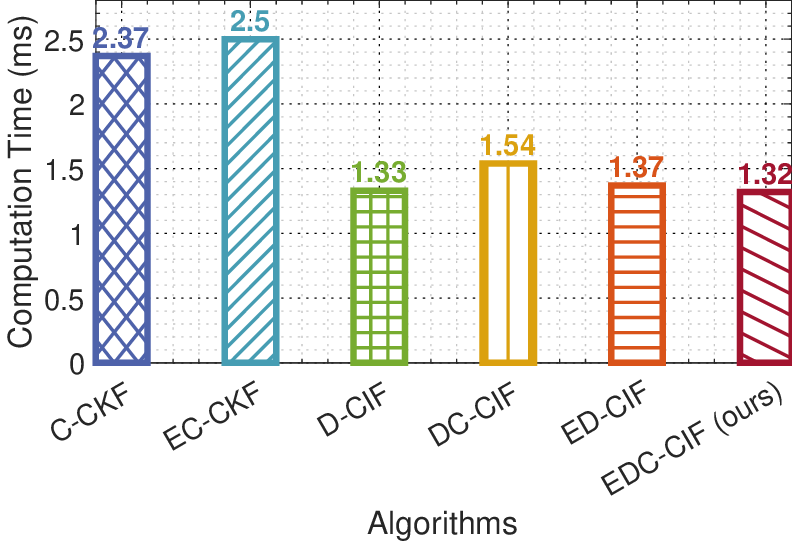}}
	\subfigure[Sensor node number $N=8$]{\includegraphics[scale=0.4]{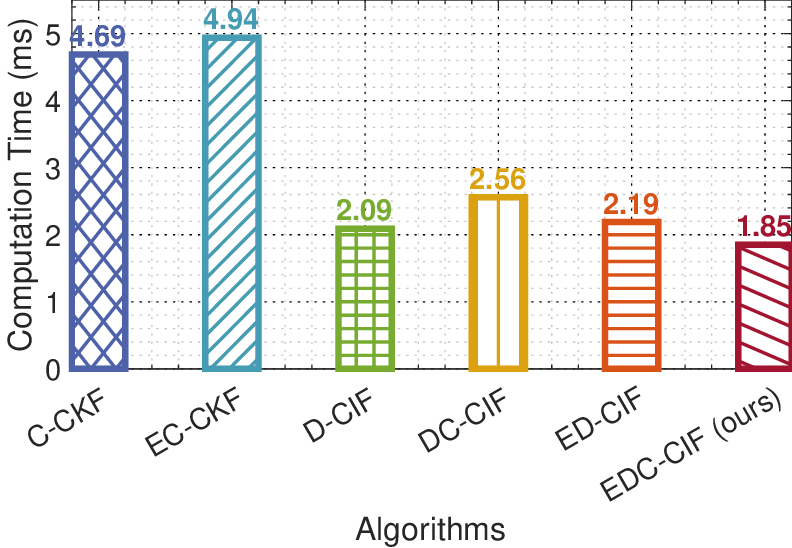}}
	\caption{Computation time of distributed filters with different node numbers.}
	\label{computation time}
 \vspace{-0.4cm}
\end{figure}


\vspace{-0.3cm}
\subsection{Communication Efficiency}
As illustrated in Fig.~\ref{TT}, the sparsity of each sensor's trigger instants through the first 100 time instants demonstrates the algorithm's effectiveness in reducing sensor nodes' communication frequency and energy consumption. 
To quantify the system's communication efficiency, we define the average trigger rate as
\begin{equation*}
    TR=\frac{1}{N}\sum_{i=1}^{N}TR_i=\frac{1}{N}\sum_{i=1}^{N}\frac{tr_i}{T}\times 100 \%,
    \label{eq TR}
\end{equation*}
where $tr_i$ denotes node $\nu_i$'s total number of trigger instants. {The system achieves an average trigger rate of 52.31\% over the entire simulation when the trigger threshold is set to 0.04, reducing the communication frequency by 47.69\% on average.} Each node's trigger rate throughout the simulation is also given in Fig.~\ref{TT}.
Since all sensors' trigger rates are smaller than $1$, the system's communication consumption is effectively reduced. {In real-world experiments, the data transmission frequency is also effectively reduced with an average trigger rate of $71.69\%$, demonstrating that our communication-efficient algorithm not only ensures tracking accuracy but also effectively reduces communication overhead. Therefore, EDC-CIF is well-suited for real-world UAV target tracking with limited bandwidth and energy constraints.}
 \vspace{-0.35cm}

\begin{figure}[htbp]
	\centering
	\includegraphics[scale=0.33]{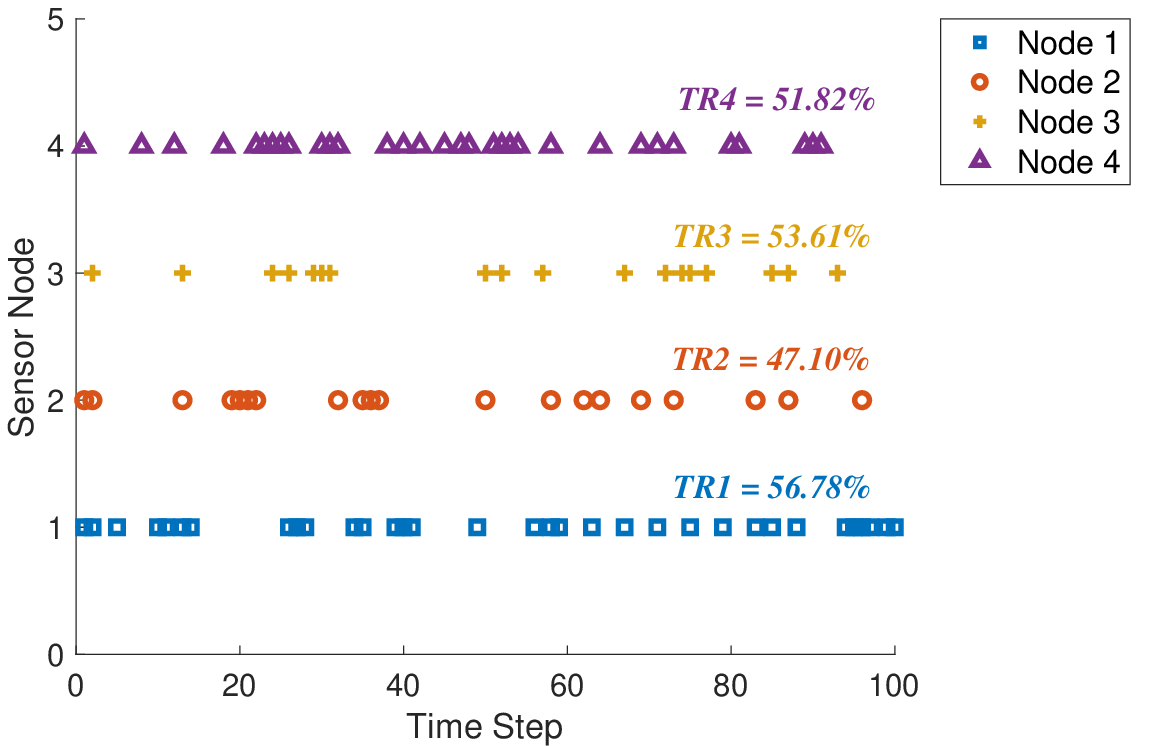}
	\caption{Trigger instants of EDC-CIF.}
	\label{TT}
 \vspace{-0.7cm}
\end{figure}



\subsection{Ablation Study} \label{ablation}
In this section, we consider two parameters' impacts on the performance of EDC-CIF. { For additional ablation studies on the two primarily designed
modules and different ET mechanisms, please refer to Appendix F in the supplementary material}.
\begin{itemize}

    \item \textbf{Trigger threshold.} \textcolor{black}{The trigger threshold $\delta$ plays a key role in the event-triggered mechanism by enabling communication-efficient filtering and regulating the transmission frequency of sensor measurements. In particular, it directly determines the degree of communication resource saving and can be selected according to the available communication bandwidth and the filtering accuracy requirements of practical applications. With different predefined trigger thresholds $\delta$ (where $\delta^i=\delta$ for $\forall \nu_i\in\boldsymbol{\mathcal{V}}$), the average RMSE and trigger rate of EDC-CIF are shown in Fig.~\ref{ab_tr}. The trigger rates of the four sensors under different thresholds are further illustrated in Fig.~\ref{TT}. As observed, the estimation error increases as the trigger threshold becomes larger, while the trigger rate correspondingly decreases. This behavior reveals the intrinsic trade-off between communication frequency and estimation accuracy. Based on the experimental results, a threshold in the range of $0.03$–$0.05$ achieves a favorable balance between communication cost and tracking performance in the simulation scenario.}
    

\begin{figure}[htbp]
	\centering
	\includegraphics[scale=0.35]{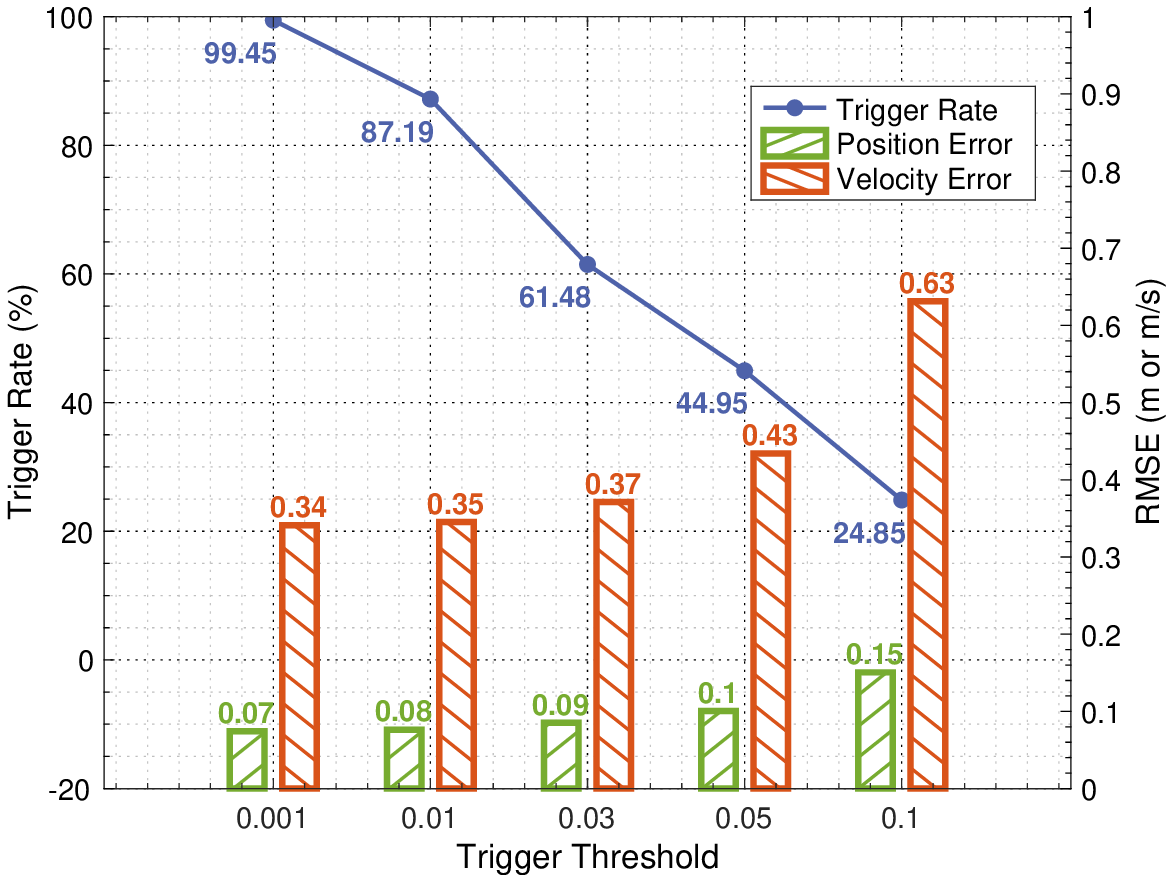}
	\caption{\textcolor{black}{Average trigger rate (line plot, left y-axis) and average RMSE of position and velocity (bar plots, right y-axis) under different trigger thresholds.}}
	\label{ab_tr}
 \vspace{-0.2cm}
\end{figure}




    \item \textbf{Scaling parameter.} 
    \textcolor{black}{
    The scaling parameters $\sigma_1$ and $\sigma_2$ are positive constants used to adjust the conservativeness of the approximated upper bound of the error covariance matrix. By scaling the estimation uncertainty, $\sigma_1$ and $\sigma_2$ play a critical role in balancing the estimation accuracy and robustness in the filter update process.
    To analyze their individual effects, a control variates method is adopted to evaluate the influence of $\sigma_1$ and $\sigma_2$ on tracking accuracy. The results are summarized in Table~\ref{sigma}, where the non-varied parameter is fixed at $5\times10^{-4}$.
    From the average RMSE values under different parameter settings, it can be observed that the tracking accuracy decreases as $\sigma_1$ increases, whereas the algorithm is largely insensitive to $\sigma_2$ over a wide range. This robustness significantly reduces the need for manual tuning of $\sigma_2$. Based on these results, both parameters are set to $5\times10^{-4}$ in the simulations to ensure stable tracking performance.
    }

    \setlength{\tabcolsep}{2.2mm}
\begin{table}[H]
	\caption{Average RMSE of EDC-CIF with different scaling parameters.}
	\begin{tabular}{ccccc}
		\toprule
		Scaling Parameter $\sigma_1$ & $5\times10^{-4}$ & $5\times10^{-3}$ & $0.05$ & $0.1$\\
		\midrule
		RMSE$^{\text{pos}}$ (m) & $0.0906$ & $0.0917$ & $0.1298$ & $0.1717$ \\  [2pt]
		RMSE$^{\text{vel}}$ (m/s) & $0.3969$ & $0.3993$ & $0.5963$ & $1.1445$ \\  [2pt]
            \toprule
            Scaling Parameter $\sigma_2$ & $5\times10^{-4}$ & $5\times10^{-3}$ & $0.05$ & $0.1$\\
		\midrule
		RMSE$^{\text{pos}}$ (m) & $0.0906$ & $0.0905$ & $0.0905$ & $0.0905$ \\  [2pt]
		RMSE$^{\text{vel}}$ (m/s) & $0.3969$ & $0.3969$ & $0.3969$ & $0.3969$ \\  [2pt]
		\bottomrule
	\end{tabular}
 \label{sigma}
 \vspace{-0.8em} 
\end{table}
    



\end{itemize}

\section{Conclusion} \label{conclusion}

This paper proposes a distributed nonlinear state estimation method for fast and accurate multi-agent collaborative target tracking with low data transmission.
The developed collaborative filter applies the ET mechanism and diffusion strategy with CI to CIF by designing an error-minimized local filter and a global correlation-aware diffusion fusion strategy. 
Simulation and real-world evaluation results show that our approach can reduce communication overhead and computation time while providing more accurate estimation for collaborative tracking compared with conventional consensus-based schemes.
\textcolor{black}{Future work will explore more adaptive strategies and learning-based methods to further improve communication efficiency.}



\section*{Acknowledgment}
This paper was supported by the Natural Science Foundation of China under Grant 62371269, Shenzhen Science and Technology Program (Grant No.  ZDCYKCX20250901094203005 and ZDCY202517014).

\bibliography{References.bib}

@inproceedings{CI,
  title={A non-divergent estimation algorithm in the presence of unknown correlations},
  author={Julier, Simon J and Uhlmann, Jeffrey K},
  booktitle={Proceedings of the 1997 American Control Conference},
  volume={4},
  pages={2369--2373},
  year={1997}
}

@article{song2020distributed,
	title={Distributed auxiliary particle filtering with diffusion strategy for target tracking: A dynamic event-triggered approach},
	author={Song, Weihao and Wang, Zidong and others},
	journal={IEEE Transactions on Signal Processing},
	volume={69},
	pages={328--340},
	year={2020}
}

@inproceedings{CCKF,
  title={Weighted average consensus-based cubature {Kalman} filtering for mobile sensor networks with switching topologies},
  author={Tan, Qingke and Dong, Xiwang and Li, Qingdong and Ren, Zhang},
  booktitle={2017 13th IEEE International Conference on Control \& Automation (ICCA)},
  pages={271--276},
  year={2017}
}

@article{SECCKF,
  title={Application of event-triggered cubature {Kalman} filter for remote nonlinear state estimation in wireless sensor network},
  author={Li, Sen and Li, Zhen and Li, Jian and others},
  journal={IEEE Transactions on Industrial Electronics},
  volume={68},
  number={6},
  pages={5133--5145},
  year={2020}
}

@article{SECEKF,
  title={Event-Triggered Extended {Kalman} Filtering Analysis for Networked Systems},
  author={Zhao, Huijuan and others},
  journal={Mathematics},
  volume={10},
  number={6},
  pages={927},
  year={2022}
}

@article{SECUKF,
  title={Event-triggered cooperative unscented {Kalman} filtering and its application in {multi-UAV} systems},
  author={Song, Weihao and Wang, Jianan and Zhao, Shiyu and Shan, Jiayuan},
  journal={Automatica},
  volume={105},
  pages={264--273},
  year={2019}
}

@article{kwon2018sensing,
  title={Sensing-based distributed state estimation for cooperative multiagent systems},
  author={Kwon, Cheolhyeon and Hwang, Inseok},
  journal={IEEE Transactions on Automatic Control},
  volume={64},
  number={6},
  pages={2368--2382},
  year={2018}
}

@article{zha2023privacy,
  title={Privacy-preserving push-sum distributed cubature information filter for nonlinear target tracking with switching directed topologies},
  author={Zha, Jirong and Han, Liang and Dong, Xiwang and others},
  journal={ISA Transactions},
  volume={136},
  pages={16--30},
  year={2023}
}

@article{degroot1974reaching,
  title={Reaching a consensus},
  author={DeGroot, Morris H},
  journal={Journal of the American Statistical Association},
  volume={69},
  number={345},
  pages={118--121},
  year={1974}
}

@article{tu2012diffusion,
  title={Diffusion strategies outperform consensus strategies for distributed estimation over adaptive networks},
  author={Tu, Sheng-Yuan and others},
  journal={IEEE Transactions on Signal Processing},
  volume={60},
  number={12},
  pages={6217--6234},
  year={2012}
}

@article{chen2021distributed,
  title={Distributed diffusion unscented {Kalman} filtering based on covariance intersection with intermittent measurements},
  author={Chen, Hao and Wang, Jianan and Wang, Chunyan and Shan, Jiayuan and Xin, Ming},
  journal={Automatica},
  volume={132},
  pages={109769},
  year={2021}
}

@inproceedings{cattivelli2010distributed,
  title={Distributed nonlinear {Kalman} filtering with applications to wireless localization},
  author={Cattivelli, Federico S and Sayed, Ali H},
  booktitle={2010 IEEE International Conference on Acoustics, Speech and Signal Processing},
  pages={3522--3525},
  year={2010}
}

@article{cattivelli2010diffusion,
  title={Diffusion strategies for distributed {Kalman} filtering and smoothing},
  author={Cattivelli, Federico S and Sayed, Ali H},
  journal={IEEE Transactions on Automatic Control},
  volume={55},
  number={9},
  pages={2069--2084},
  year={2010}
}

@article{vahidpour2019partial,
  title={Partial diffusion {Kalman} filtering for distributed state estimation in multiagent networks},
  author={Vahidpour, Vahid and Rastegarnia, Amir and others},
  journal={IEEE Transactions on Neural Networks and Learning Systems},
  volume={30},
  number={12},
  pages={3839--3846},
  year={2019}
}

@article{liang2023event,
  title={Event-triggered diffusion nonlinear estimation for sensor networks with unknown cross-correlations},
  author={Liang, Yuan and Li, Yinya and Chen, Ye and Sheng, Andong},
  journal={Systems \& Control Letters},
  volume={175},
  pages={105506},
  year={2023}
}

@article{gao2023robust,
  title={Robust integrated sequential covariance intersection fusion {Kalman} filters and their convergence and stability for networked sensor systems with five uncertainties},
  author={Gao, Yuan and Deng, Zili},
  journal={International Journal of Robust and Nonlinear Control},
  volume={33},
  number={2},
  pages={1371--1406},
  year={2023}
}

@article{hu2011diffusion,
  title={Diffusion {Kalman} filtering based on covariance intersection},
  author={Hu, Jinwen and Xie, Lihua and Zhang, Cishen},
  journal={IEEE Transactions on Signal Processing},
  volume={60},
  number={2},
  pages={891--902},
  year={2011}
}

@article{sun2023inverse,
  title={Inverse-Covariance-Intersection-Based Distributed Estimation and Application in Wireless Sensor Network},
  author={Sun, Tao and others},
  journal={IEEE Transactions on Industrial Informatics},
  volume={19},
  number={10},
  pages={10079-10090},
  year={2023}
}

@article{hu2023distributed,
  title={Distributed resilient fusion filtering for nonlinear systems with multiple missing measurements via dynamic event-triggered mechanism},
  author={Hu, Jun and Hu, Zhibin and others},
  journal={Information Sciences},
  volume={637},
  pages={118950},
  year={2023}
}

@article{peng2018survey,
  title={A survey on recent advances in event-triggered communication and control},
  author={Peng, Chen and others},
  journal={Information Sciences},
  volume={457},
  pages={113--125},
  year={2018}
}

@article{miskowicz2006send,
  title={Send-on-delta concept: An event-based data reporting strategy},
  author={Miskowicz, Marek},
  journal={Sensors},
  volume={6},
  number={1},
  pages={49--63},
  year={2006}
}

@article{li2016event,
  title={Event-triggered {Kalman} consensus filter over sensor networks},
  author={Li, Wenling and Jia, Yingmin and Du, Junping},
  journal={IET Control Theory \& Applications},
  volume={10},
  number={1},
  pages={103--110},
  year={2016}
}

@article{zhang2017distributed,
  title={Distributed {Kalman} consensus filter with event-triggered communication: Formulation and stability analysis},
  author={Zhang, Cui and Jia, Yingmin},
  journal={Journal of the Franklin Institute},
  volume={354},
  number={13},
  pages={5486--5502},
  year={2017}
}

@article{ge2021dynamic,
  title={Dynamic event-triggered control and estimation: A survey},
  author={Ge, Xiaohua and Han, Qing-Long and Zhang, Xian-Ming and Ding, Derui},
  journal={International Journal of Automation and Computing},
  volume={18},
  number={6},
  pages={857--886},
  year={2021}
}

@article{rezaei2022event,
  title={Event-triggered resilient distributed extended {Kalman} filter with consensus on estimation},
  author={Rezaei, Hossein and others},
  journal={International Journal of Robust and Nonlinear Control},
  volume={32},
  number={3},
  pages={1303--1315},
  year={2022}
}

@article{battistelli2018distributed,
  title={A distributed {Kalman} filter with event-triggered communication and guaranteed stability},
  author={Battistelli, Giorgio and Chisci, Luigi and Selvi, Daniela},
  journal={Automatica},
  volume={93},
  pages={75--82},
  year={2018}
}

@article{tan2018distributed,
  title={Distributed event-triggered cubature information filtering based on weighted average consensus},
  author={Tan, Qingke and Dong, Xiwang and Li, Qingdong and Ren, Zhang},
  journal={IET Control Theory \& Applications},
  volume={12},
  number={1},
  pages={78--86},
  year={2018}
}

@inproceedings{lin2022distributed,
  title={Distributed cubature information filtering based on hybrid consensus strategy with event-triggered mechanism},
  author={Lin, Xufeng and Zhang, Xuechun and others},
  booktitle={2022 13th Asian Control Conference (ASCC)},
  pages={2138--2143},
  year={2022}
}

@article{han2015stochastic,
  title={Stochastic event-triggered sensor schedule for remote state estimation},
  author={Han, Duo and Mo, Yilin and others},
  journal={IEEE Transactions on Automatic Control},
  volume={60},
  number={10},
  pages={2661--2675},
  year={2015}
}

@article{hu2015consensus,
  title={Consensus of linear multi-agent systems by distributed event-triggered strategy},
  author={Hu, Wenfeng and Liu, Lu and Feng, Gang},
  journal={IEEE Transactions on Cybernetics},
  volume={46},
  number={1},
  pages={148--157},
  year={2015}
}

@article{kamal2013information,
  title={Information weighted consensus filters and their application in distributed camera networks},
  author={Kamal, Ahmed T and others},
  journal={IEEE Transactions on Automatic Control},
  volume={58},
  number={12},
  pages={3112--3125},
  year={2013}
}

@article{cao2012overview,
  title={An overview of recent progress in the study of distributed multi-agent coordination},
  author={Cao, Yongcan and Yu, Wenwu and Ren, Wei and Chen, Guanrong},
  journal={IEEE Transactions on Industrial informatics},
  volume={9},
  number={1},
  pages={427--438},
  year={2012}
}

@article{arasaratnam2009cubature,
  title={Cubature {Kalman} filters},
  author={Arasaratnam, Ienkaran and others},
  journal={IEEE Transactions on Automatic Control},
  volume={54},
  number={6},
  pages={1254--1269},
  year={2009}
}

@inproceedings{pakki2011cubature,
  title={Cubature information filter and its applications},
  author={Pakki, Kumar and Chandra, Bharani and Gu, Da-Wei and Postlethwaite, Ian},
  booktitle={Proceedings of the 2011 American Control Conference},
  pages={3609--3614},
  year={2011}
}

@article{marano2008distributed,
  title={Distributed estimation in large wireless sensor networks via a locally optimum approach},
  author={Marano, Stefano and Matta, Vincenzo and Willett, Peter},
  journal={IEEE Transactions on Signal Processing},
  volume={56},
  number={2},
  pages={748--756},
  year={2008}
}

@article{marano2006quantizer,
  title={Quantizer precision for distributed estimation in a large sensor network},
  author={Marano, Stefano and Matta, Vincenzo and others},
  journal={IEEE Transactions on Signal Processing},
  volume={54},
  number={10},
  pages={4073--4078},
  year={2006}
}

@inproceedings{liu2017multi,
  title={Multi-agent formation control with target tracking and navigation},
  author={Liu, Lebin and Luo, Chaomin and Shen, Furao},
  booktitle={2017 IEEE International Conference on Information and Automation (ICIA)},
  pages={98--103},
  year={2017}
}

@article{wang2018diffusion,
  title={Diffusion nonlinear Kalman filter with intermittent observations},
  author={Wang, Guoqing and Li, Ning and Zhang, Yonggang},
  journal={Proceedings of the Institution of Mechanical Engineers, Part G: Journal of Aerospace Engineering},
  volume={232},
  number={15},
  pages={2775--2783},
  year={2018},
  publisher={SAGE Publications Sage UK: London, England}
}

@article{win2011network,
  title={Network localization and navigation via cooperation},
  author={Win, Moe Z and Conti, Andrea and Mazuelas, Santiago and others},
  journal={IEEE Communications Magazine},
  volume={49},
  number={5},
  pages={56--62},
  year={2011},
  publisher={IEEE}
}

@misc{grupp2017evo,
  title={{EVO: Python package for the evaluation of odometry and SLAM}},
  author={Grupp, Michael},
  year={2017}
}

@inproceedings{sturm2012benchmark,
  title={{A benchmark for the evaluation of RGB-D SLAM systems}},
  author={Sturm, J{\"u}rgen and Engelhard, Nikolas and others},
  booktitle={2012 IEEE/RSJ International Conference on Intelligent Robots and Systems},
  pages={573--580},
  year={2012},
  organization={IEEE}
}

@article{gu2025mr,
  title={{MR-COGraphs: Communication-efficient Multi-Robot Open-vocabulary Mapping System via 3D Scene Graphs}},
  author={Gu, Qiuyi and Ye, Zhaocheng and Yu, Jincheng and others},
  journal={IEEE Robotics and Automation Letters},
  year={2025},
  publisher={IEEE}
}

@inproceedings{cheng2024multi,
  title={Multi-Agent Target Pursuit Using Perception Uncertainty-Aware Reinforcement Learning},
  author={Cheng, Yuhan and Zha, Jirong and Yang, Renjue and others},
  booktitle={Proceedings of the 30th Annual International Conference on Mobile Computing and Networking},
  pages={1992--1997},
  year={2024}
}

@inproceedings{zha2026dimm,
  title={DIMM: Decoupled Multi-hierarchy Kalman Filter via Reinforcement Learning},
  author={Zha, Jirong and Fan, Yuxuan and Li, Kai and Li, Han and Gao, Chen and Chen, Xinlei},
  booktitle={Proceedings of the AAAI Conference on Artificial Intelligence},
  volume={40},
  number={22},
  pages={18746--18754},
  year={2026}
}

@article{wang2025aerial,
  title={Aerial Shepherds: Enabling Hierarchical Localization in Heterogeneous MAV Swarms},
  author={Wang, Haoyang and Xu, Jingao and Zhao, Chenyu and others},
  journal={arXiv preprint arXiv:2506.08408},
  year={2025}
}

@inproceedings{zha2026aircopbench,
  title={Aircopbench: A benchmark for multi-drone collaborative embodied perception and reasoning},
  author={Zha, Jirong and Fan, Yuxuan and Zhang, Tianyu and Chen, Geng and Chen, Yingfeng and Gao, Chen and Chen, Xinlei},
  booktitle={Proceedings of the AAAI Conference on Artificial Intelligence},
  volume={40},
  number={2},
  pages={1507--1515},
  year={2026}
}

@inproceedings{wang2024transformloc,
  title={Transformloc: Transforming mavs into mobile localization infrastructures in heterogeneous swarms},
  author={Wang, Haoyang and Xu, Jingao and others},
  booktitle={IEEE INFOCOM 2024-IEEE Conference on Computer Communications},
  pages={1101--1110},
  year={2024},
  organization={IEEE}
}

@article{dai2021adaptive,
  title={Adaptive finite-time tracking control of nonholonomic multirobot formation systems with limited field-of-view sensors},
  author={Dai, Shi-Lu and Lu, Ke and Fu, Jun},
  journal={IEEE Transactions on cybernetics},
  volume={52},
  number={10},
  pages={10695--10708},
  year={2021},
  publisher={IEEE}
}

@article{zheng2025optimal,
  title={Optimal spatial--temporal triangulation for bearing-only cooperative motion estimation},
  author={Zheng, Canlun and Mi, Yize and Guo, Hanqing and Chen, Huaben and Lin, Zhiyun and Zhao, Shiyu},
  journal={Automatica},
  volume={175},
  pages={112216},
  year={2025},
  publisher={Elsevier}
}

@article{zhou2022integrated,
  title={Integrated sensing and communication in UAV swarms for cooperative multiple targets tracking},
  author={Zhou, Longyu and Leng, Supeng and Wang, Qing and Liu, Qiang},
  journal={IEEE Transactions on Mobile Computing},
  volume={22},
  number={11},
  pages={6526--6542},
  year={2022},
  publisher={IEEE}
}

@article{gu2018multiple,
  title={Multiple moving targets surveillance based on a cooperative network for multi-UAV},
  author={Gu, Jingjing and Su, Tao and Wang, Qiuhong and Du, Xiaojiang and Guizani, Mohsen},
  journal={IEEE Communications Magazine},
  volume={56},
  number={4},
  pages={82--89},
  year={2018},
  publisher={IEEE}
}

@inproceedings{olfati2007distributed,
  title={Distributed Kalman filtering for sensor networks},
  author={Olfati-Saber, Reza},
  booktitle={2007 46th IEEE conference on decision and control},
  pages={5492--5498},
  year={2007},
  organization={IEEE}
}

@article{zhu2020multi,
  title={Multi-drone-based single object tracking with agent sharing network},
  author={Zhu, Pengfei and Zheng, Jiayu and Du, Dawei and Wen, Longyin and Sun, Yiming and Hu, Qinghua},
  journal={IEEE Transactions on Circuits and Systems for Video Technology},
  volume={31},
  number={10},
  pages={4058--4070},
  year={2020},
  publisher={IEEE}
}

@article{li2022resilient,
  title={Resilient unscented Kalman filtering fusion with dynamic event-triggered scheme: Applications to multiple unmanned aerial vehicles},
  author={Li, Chunyu and Wang, Zidong and Song, Weihao and Zhao, Shixin and Wang, Jianan and Shan, Jiayuan},
  journal={IEEE Transactions on Control Systems Technology},
  volume={31},
  number={1},
  pages={370--381},
  year={2022},
  publisher={IEEE}
}

@article{tian2024ucdnet,
  title={Ucdnet: Multi-uav collaborative 3-d object detection network by reliable feature mapping},
  author={Tian, Pengju and Wang, Zhirui and Cheng, Peirui and Wang, Yuchao and Wang, Zhechao and Zhao, Liangjin and Yan, Menglong and Yang, Xue and Sun, Xian},
  journal={IEEE Transactions on Geoscience and Remote Sensing},
  volume={63},
  pages={1--16},
  year={2024},
  publisher={IEEE}
}

@article{hu2022where2comm,
  title={Where2comm: Communication-efficient collaborative perception via spatial confidence maps},
  author={Hu, Yue and Fang, Shaoheng and Lei, Zixing and Zhong, Yiqi and Chen, Siheng},
  journal={Advances in neural information processing systems},
  volume={35},
  pages={4874--4886},
  year={2022}
}

@article{xu2026scalable,
  title={Scalable UAV multi-hop networking via multi-agent reinforcement learning with large language models},
  author={Xu, Yanggang and Zha, Jirong and Hong, Weijie and Yi, Xiangmin and Chen, Geng and Zheng, Jianfeng and Hsia, Chen-Chun and Chen, Xinlei},
  journal={IEEE Transactions on Mobile Computing},
  year={2026},
  publisher={IEEE}
}

@article{wang2026event,
  title={Event camera meets mobile embodied perception: abstraction, algorithm, acceleration, application},
  author={Wang, Haoyang and Guo, Ruishan and Ma, Pengtao and Ruan, Ciyu and Luo, Xinyu and Ding, Wenhua and Zhong, Tianyang and Xu, Jingao and Liu, Yunhao and Chen, Xinlei},
  journal={ACM Computing Surveys},
  volume={58},
  number={8},
  pages={1--41},
  year={2026},
  publisher={ACM New York, NY}
}

@inproceedings{wang2025ultra,
  title={Ultra-high-frequency harmony: mmwave radar and event camera orchestrate accurate drone landing},
  author={Wang, Haoyang and Xu, Jingao and Luo, Xinyu and Chen, Xuecheng and Zhang, Ting and Duan, Ruiyang and Liu, Yunhao and Chen, Xinlei},
  booktitle={Proceedings of the 23rd ACM Conference on Embedded Networked Sensor Systems},
  pages={15--29},
  year={2025}
}

@article{abdulghafoor2023multi,
  title={Multi-agent distributed optimal control for tracking large-scale multi-target systems in dynamic environments},
  author={Abdulghafoor, Alaa Z and Bakolas, Efstathios},
  journal={IEEE Transactions on Cybernetics},
  volume={54},
  number={5},
  pages={2866--2879},
  year={2023},
  publisher={IEEE}
}

@article{peng2025multi,
  title={Multi-UAV cooperative pursuit strategy with limited visual field in urban airspace: A multi-agent reinforcement learning approach},
  author={Peng, Zhe and Wu, Guohua and Luo, Biao and Wang, Ling},
  journal={IEEE/CAA Journal of Automatica Sinica},
  year={2025},
  publisher={IEEE}
}

@article{hou2023uav,
  title={UAV swarm cooperative target search: A multi-agent reinforcement learning approach},
  author={Hou, Yukai and Zhao, Jin and Zhang, Rongqing and Cheng, Xiang and Yang, Liuqing},
  journal={IEEE Transactions on Intelligent Vehicles},
  volume={9},
  number={1},
  pages={568--578},
  year={2023},
  publisher={IEEE}
}

@article{wan2024deep,
  title={Deep reinforcement learning enabled multi-UAV scheduling for disaster data collection with time-varying value},
  author={Wan, Pengfu and Xu, Gangyan and Chen, Jiawei and Zhou, Yaoming},
  journal={IEEE Transactions on Intelligent Transportation Systems},
  volume={25},
  number={7},
  pages={6691--6702},
  year={2024},
  publisher={IEEE}
}

@article{liu2022smart,
  title={SMART: Vision-based method of cooperative surveillance and tracking by multiple UAVs in the urban environment},
  author={Liu, Daqian and Zhu, Xiaomin and Bao, Weidong and Fei, Bowen and Wu, Jianhong},
  journal={IEEE Transactions on Intelligent Transportation Systems},
  volume={23},
  number={12},
  pages={24941--24956},
  year={2022},
  publisher={IEEE}
}

@article{javaid2023communication,
  title={Communication and control in collaborative UAVs: Recent advances and future trends},
  author={Javaid, Shumaila and Saeed, Nasir and Qadir, Zakria and Fahim, Hamza and He, Bin and Song, Houbing and Bilal, Muhammad},
  journal={IEEE Transactions on Intelligent Transportation Systems},
  volume={24},
  number={6},
  pages={5719--5739},
  year={2023},
  publisher={IEEE}
}

@article{miao2025multi,
  title={Multi-UAV Cooperative Target Consensus Detection and Trajectory Optimization in Urban Environments},
  author={Miao, Kunzhong and Wang, Chang and Niu, Yifeng and Yu, Huangzhi and Liu, Tianqing},
  journal={IEEE Transactions on Aerospace and Electronic Systems},
  year={2025},
  publisher={IEEE}
}

@article{li2025distributed,
  title={A distributed cooperative dynamic target search method for multi-UAV systems in complex adversarial environments},
  author={Li, Yiyuan and Chen, Weiyi and Fu, Bing and Liu, Shukan and Hao, Lingjun and Wu, Zhonghong},
  journal={IEEE Internet of Things Journal},
  year={2025},
  publisher={IEEE}
}

@article{beard2006decentralized,
  title={Decentralized cooperative aerial surveillance using fixed-wing miniature UAVs},
  author={Beard, Randal W and McLain, Timothy W and Nelson, Derek B and Kingston, Derek and Johanson, David},
  journal={Proceedings of the IEEE},
  volume={94},
  number={7},
  pages={1306--1324},
  year={2006},
  publisher={IEEE}
}

@article{zhou2021intelligent,
  title={Intelligent UAV swarm cooperation for multiple targets tracking},
  author={Zhou, Longyu and Leng, Supeng and Liu, Qiang and Wang, Qing},
  journal={IEEE Internet of Things Journal},
  volume={9},
  number={1},
  pages={743--754},
  year={2021},
  publisher={IEEE}
}

@article{wang2025uav,
  title={UAV--Ground vehicle collaborative delivery in emergency response: A review of key technologies and future trends},
  author={Wang, Yizhe and Li, Jie and Yang, Xiaoguang and Peng, Qing},
  journal={Applied Sciences},
  volume={15},
  number={17},
  pages={9803},
  year={2025},
  publisher={MDPI}
}

@article{khan2022emerging,
  title={Emerging UAV technology for disaster detection, mitigation, response, and preparedness},
  author={Khan, Amina and Gupta, Sumeet and Gupta, Sachin Kumar},
  journal={Journal of Field Robotics},
  volume={39},
  number={6},
  pages={905--955},
  year={2022},
  publisher={Wiley Online Library}
}

@inproceedings{stasinchuk2021multi,
  title={A multi-UAV system for detection and elimination of multiple targets},
  author={Stasinchuk, Yurii and Vrba, Matou{\v{s}} and Petrl{\'\i}k, Mat{\v{e}}j and B{\'a}{\v{c}}a, Tom{\'a}{\v{s}} and Spurn{\`y}, Vojt{\v{e}}ch and Hert, Daniel and {\v{Z}}aitl{\'\i}k, David and Nascimento, Tiago and Saska, Martin},
  booktitle={2021 IEEE international conference on robotics and automation (ICRA)},
  pages={555--561},
  year={2021},
  organization={IEEE}
}

@article{zhang2023integrated,
  title={Integrated design of cooperative area coverage and target tracking with multi-UAV system},
  author={Zhang, Mengge and Wu, Xinning and Li, Jie and Wang, Xiangke and Shen, Lincheng},
  journal={Journal of Intelligent \& Robotic Systems},
  volume={108},
  number={4},
  pages={77},
  year={2023},
  publisher={Springer}
}

@article{du2025survey,
  title={A survey on autonomous and intelligent swarms of uncrewed aerial vehicles (UAVs)},
  author={Du, Zhenpeng and Luo, Chunbo and Min, Geyong and Wu, Jia and Luo, Cai and Pu, Jian and Li, Shuai},
  journal={IEEE Transactions on Intelligent Transportation Systems},
  year={2025},
  publisher={IEEE}
}

@inproceedings{peng2026resilient,
  title={Resilient UAV Swarm with Fast Connectivity Recovery and Extensive Coverage},
  author={Peng, Yabin and Zhou, Chenyu and Cui, Hainan and Duan, Tong and Chen, Haoyang and Zhang, Fan and Liu, Shaoxun},
  booktitle={Proceedings of the AAAI Conference on Artificial Intelligence},
  volume={40},
  number={2},
  pages={917--925},
  year={2026}
}

@article{zhang2025cooperative,
  title={Cooperative Dynamic Target Tracking: Distributed Time-Varying Optimization for Multi-UAV System},
  author={Zhang, Boyang and Hou, Yueqi and Yin, Hang and Lv, Maolong and Yang, Aiwu and Wu, Lirong},
  journal={IEEE Transactions on Aerospace and Electronic Systems},
  year={2025},
  publisher={IEEE}
}

@article{doostmohammadian2021distributed,
  title={Distributed estimation approach for tracking a mobile target via formation of UAVs},
  author={Doostmohammadian, Mohammadreza and Taghieh, Amin and Zarrabi, Houman},
  journal={IEEE Transactions on Automation Science and Engineering},
  volume={19},
  number={4},
  pages={3765--3776},
  year={2021},
  publisher={IEEE}
}

@inproceedings{wang2020learning,
  title={Learning efficient multi-agent communication: An information bottleneck approach},
  author={Wang, Rundong and He, Xu and Yu, Runsheng and Qiu, Wei and An, Bo and Rabinovich, Zinovi},
  booktitle={International conference on machine learning},
  pages={9908--9918},
  year={2020},
  organization={PMLR}
}

@inproceedings{berna2004communication,
  title={Communication efficiency in multi-agent systems},
  author={Berna-Koes, Mary and Nourbakhsh, Illah and Sycara, Katia},
  booktitle={IEEE International Conference on Robotics and Automation, 2004. Proceedings. ICRA'04. 2004},
  volume={3},
  pages={2129--2134},
  year={2004},
  organization={IEEE}
}

@article{sun2024efficient,
  title={Efficient joint deployment of multi-UAVs for target tracking in traffic big data},
  author={Sun, Lu and Wang, Jiashuai and Wang, Jie and Lin, Lin and Gen, Mitsuo},
  journal={IEEE Transactions on Intelligent Transportation Systems},
  volume={25},
  number={7},
  pages={7780--7791},
  year={2024},
  publisher={IEEE}
}

\vspace{-0.5cm}
\begin{IEEEbiography}
[{\includegraphics[width=1in,height=1.25in,clip,keepaspectratio]{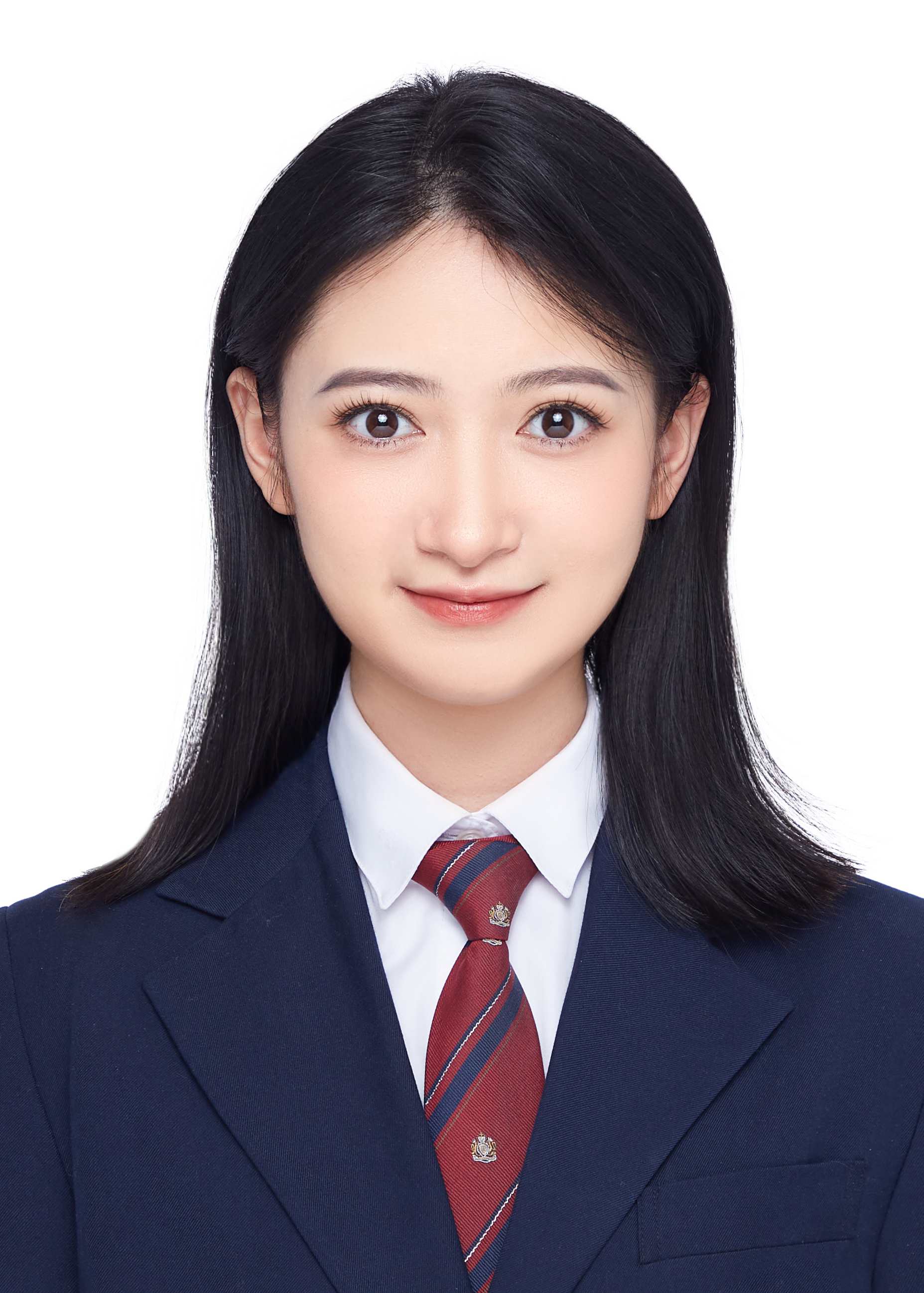}}]
{Jirong Zha}
is currently pursuing her Ph.D. degree in data science and information technology from Tsinghua University, China. She
received the M.S. and B.S. degrees from Beihang University, China, in 2023 and 2020, respectively. Her research interests include collaborative perception, large language model, distributed state estimation, and multi-agent systems.
\end{IEEEbiography}

\begin{IEEEbiography}
[{\includegraphics[width=1in,height=1.25in,clip,keepaspectratio]{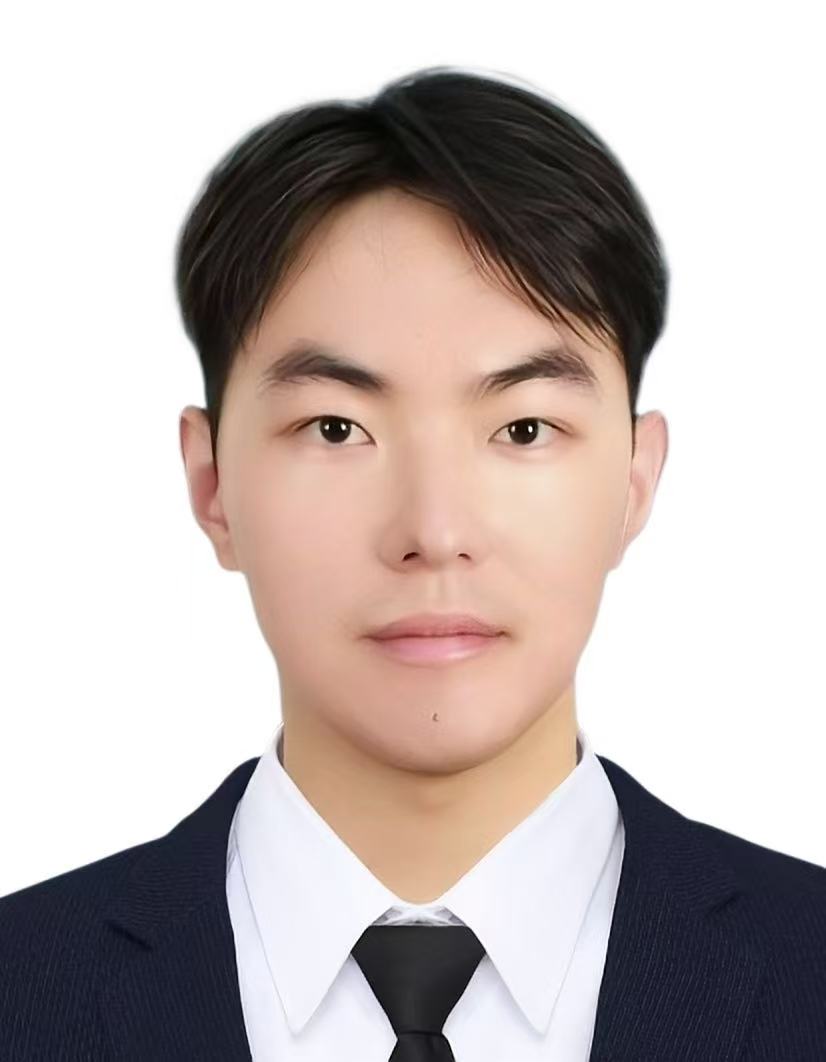}}]
{Chenyu Zhao}
is currently pursuing his Ph.D. degree in the Department of Computing, Imperial College London, UK. He received his M.Sc. degree in Data Science and Information Technology at Tsinghua University, China, in 2025. He received his B.E. degree in Electronic Information Science and Technology from Northwest University in China and his B.S. degree in Electronic System Engineering from the University of Essex in the UK in 2022. His research interests include Mobile Computing, Cyber-Physical Systems, Embedded AI, Robotics, and Quadrotors.
\end{IEEEbiography}

\begin{IEEEbiography}
[{\includegraphics[width=1in,height=1.25in,clip,keepaspectratio]{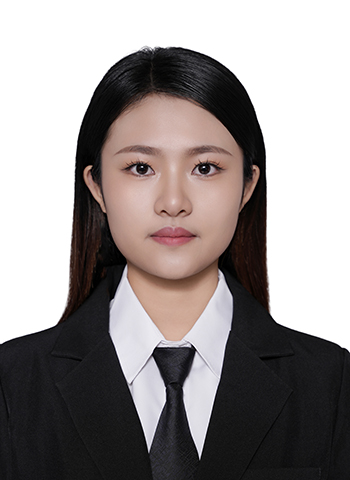}}]
{Nan Zhou}
received the B.E. degree from the School of Telecommunications Engineering, Xidian University, China, in 2024. She is currently pursuing a Master's degree at the Tsinghua Shenzhen International Graduate School, Tsinghua University, China. Her research interests include spatio-temporal dynamic, world model, and physical knowledge.
\end{IEEEbiography}

\begin{IEEEbiography}
[{\includegraphics[width=1in,height=1.25in,clip,keepaspectratio]{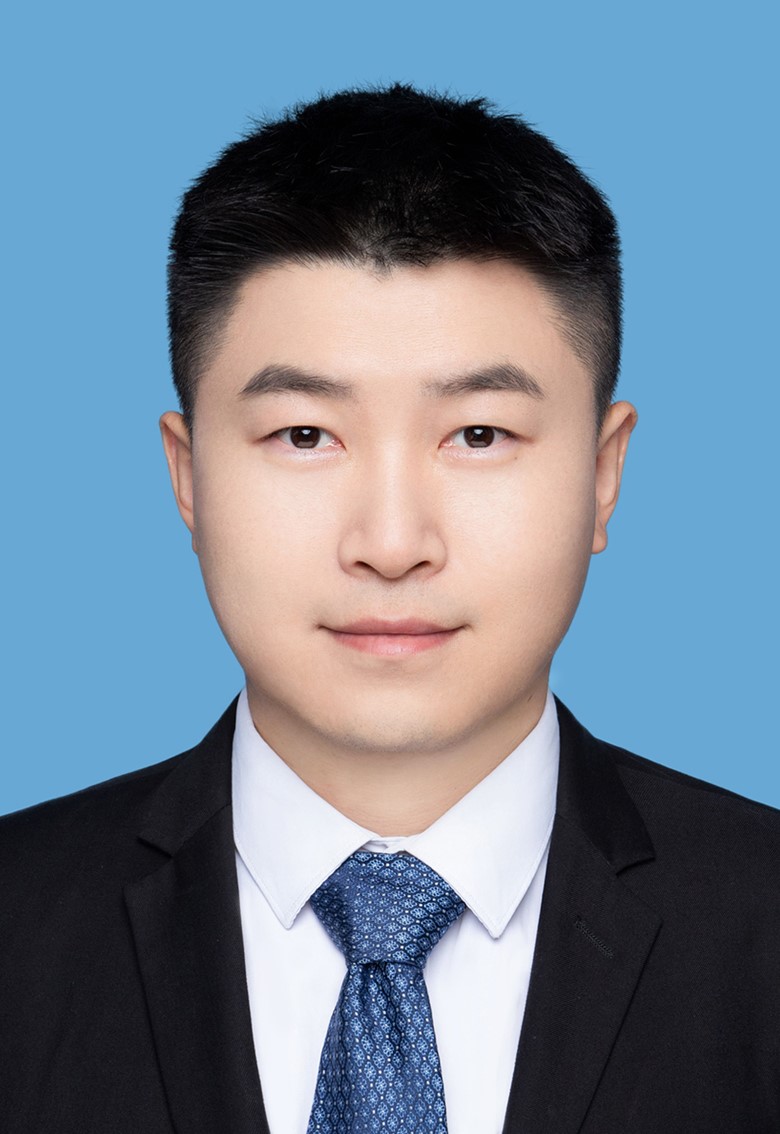}}]
{Zhenyu Liu}
is an Assistant Professor at Tsinghua Shenzhen International Graduate School, Tsinghua University, China. He received the Ph.D. degree in networks and statistics and the S.M. degree in aeronautics and astronautics both from the Massachusetts Institute of Technology (MIT) in 2022. His research interests include wireless communications, network localization, distributed inference,  networked control, and quantum information science. Dr. Liu received the Best Paper Awards at the IEEE LATINCOM in 2017 and at the WCSP in 2025. He is a Guest Editor for IEEE JSAC Quantum Series.
\end{IEEEbiography}

\begin{IEEEbiography} 
[{\includegraphics[width=1in,height=1.25in,clip,keepaspectratio]{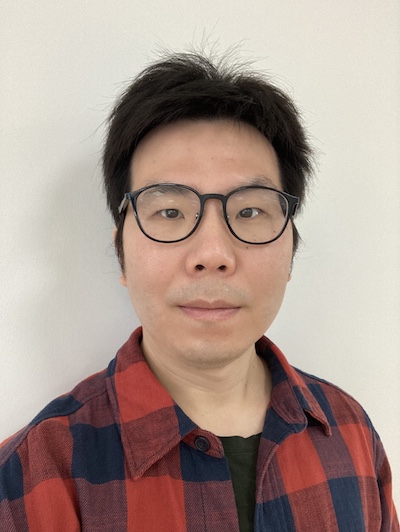}}]
{Tao Sun}
is currently working as an associate research scientist at Institute of Artificial Intelligence and Robotics for Society(AiRS) in Shenzhen, China. He received his Ph.D in Mechanical and Aerospace Engineering from University of Missouri, Columbia, USA, in 2018, M.S. in Electrical Engineering from University of California, Riverside, USA, in 2011 and B.E. in Electronics Science and Technology from Hefei University of Technology, Hefei, China, in 2009. He worked as a post-doctoral research associate at Virginia Commonwealth University, Richmond, USA during 2018-2019, R\&D Scientist in UtopiaCompression Corp., Los Angeles, USA during 2019-2021 and an assistant research fellow at Peng Cheng Laboratory, Shenzhen, China during 2021-2024. His research interests include nonlinear estimation/filtering theory, distributed estimation based on sensor network, information fusion, unmanned aerial vehicles and target tracking.
\end{IEEEbiography}

\begin{IEEEbiography} 
[{\includegraphics[width=1in,height=1.25in,clip,keepaspectratio]{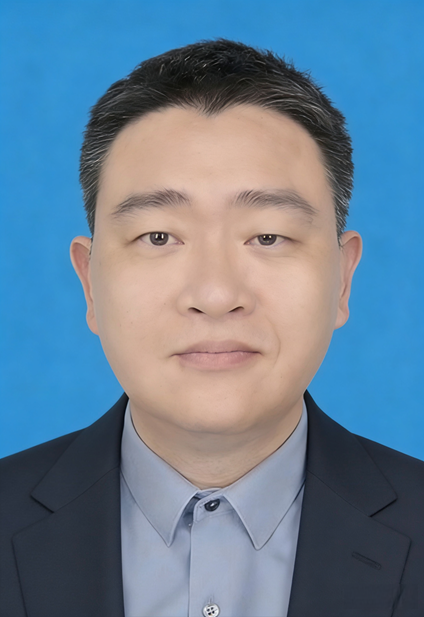}}]
{Bin Zhang}
is the Deputy Director of the Technology Department of Shenzhen Smart City Technology Development Group, and Chief Expert of the Digital Twin Special Committee of the Group Science and Technology Commission. He has long been committed to technical research in the fields of urban digital twin, spatiotemporal big data, BIM, GIS and spatiotemporal intelligence. He has presided over several city-level digital twin platform projects and major scientific and technological research projects.
\end{IEEEbiography}

\begin{IEEEbiography} 
[{\includegraphics[width=1in,height=1.25in,clip,keepaspectratio]{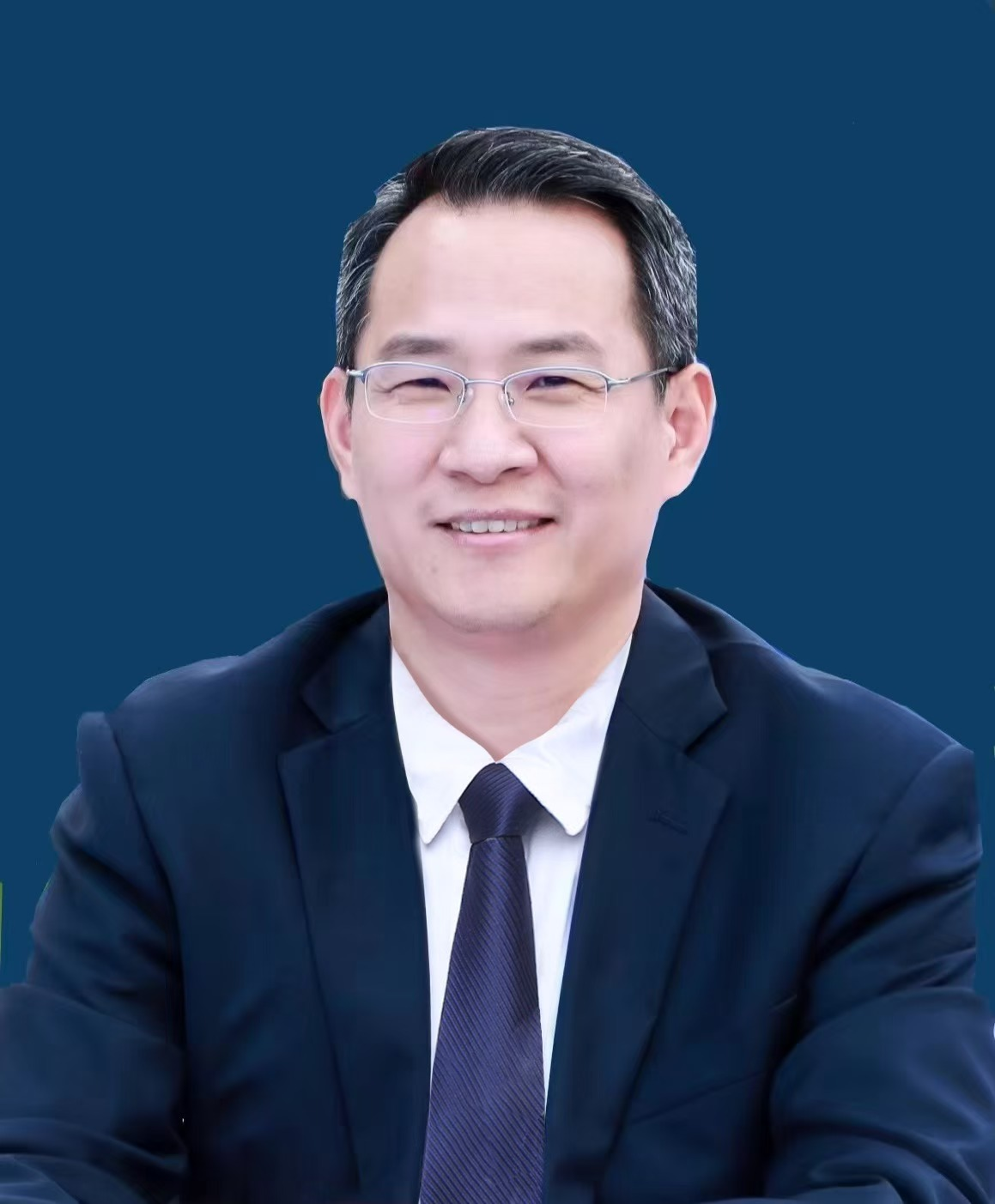}}]
{Xiaochun Zhang}
is currently the Chairman of Shenzhen Smart City Technology Development Group, holds a Doctorate in Engineering, is a professor-level senior engineer, and a national master of engineering design and geotechnique investigation. He has been devoted to research in the fields of urban digital twins, smart transportation, and the low-altitude economy for a long time, and has won multiple scientific and technological progress awards.
\end{IEEEbiography}

\begin{IEEEbiography}
[{\includegraphics[width=1in,height=1.25in,clip,keepaspectratio]{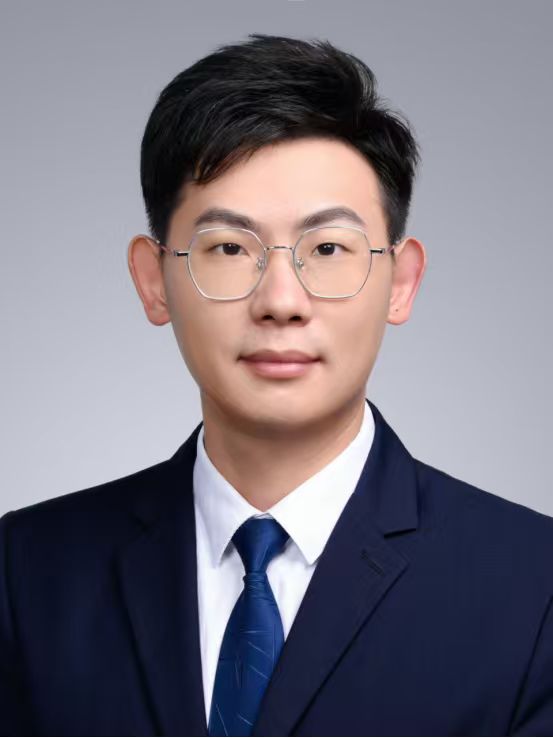}}]
{Xinlei Chen}
is an associate professor at Shenzhen International Graduate School, Tsinghua University. His research interests lie in mobile sensing, embodied AI and etc. Dr. Chen has won several awards from top-tier conference and been selected in Elsevier's Global Top 2\% Scientists List in the past three years.
\end{IEEEbiography}

\end{document}